\documentclass[symmetry,article,accept,moreauthors]{mdpi}
\usepackage{tikz}
\usetikzlibrary{arrows.meta, positioning, shapes.geometric, fit, backgrounds, calc, decorations.pathreplacing}

\usepackage{fancyhdr}
\fancypagestyle{plain}{%
  \fancyhf{}%
  \fancyhead[L]{\textit{Preprint. Work in progress.}}%
  \fancyfoot[C]{\thepage}%
}

\firstpage{1}
\pubvolume{1}
\issuenum{1}
\articlenumber{0}
\pubyear{2026}
\copyrightyear{2026}
\datereceived{ }
\daterevised{ }
\dateaccepted{ }
\datepublished{ }

\Title{A Non-Invasive Cloud-Based Migration Strategy for Post-Quantum Cybersecurity in Smart HVAC Systems: Architecture, Implementation, and Empirical Evaluation}

\Author{Mahedee Zaman Moon $^{1}$\orcidA{}, Kaysarul Anas Apurba $^{2}$\orcidB{}, Md Hasibul Hasan $^{3}$\orcidC{} and Sk Md Mizanur Rahman $^{4,}$\orcidD{}}

\AuthorNames{Mahedee Zaman Moon; Kaysarul Anas Apurba; Md Hasibul Hasan; Sk Md Mizanur Rahman}

\address{
$^{1}$ \quad Islamic University of Technology, Gazipur, Bangladesh; mahedeezaman@iut-dhaka.edu\\
$^{2}$ \quad Laurentian University, Sudbury, ON, Canada; kaysarulanas2@gmail.com\\
$^{3}$ \quad Laurentian University, Sudbury, ON, Canada; cs.hasibul@gmail.com\\
$^{4}$ \quad Centennial College, Toronto, ON, Canada;  srahman@centennialcollege.ca
}

\corres{Correspondence:  srahman@centennialcollege.ca}

\abstract{Legacy smart HVAC controllers rely on vendor-cloud TLS secured by ECDH and RSA, both broken by Shor's algorithm, and typical 10--15 year lifespans mean today's devices remain in service through the quantum-threat era. Direct on-device post-quantum cryptography is infeasible: an ESP32-S3, representative of capable HVAC hardware, has only 339 KB free heap against the 900 KB ML-KEM-768 requires, and even classical ECDH-P256 keygen (111.93 ms) dwarfs hardware AES-128 (0.032 ms). We propose a non-invasive PQC proxy, requiring no device, firmware, or vendor-cloud changes, performing ML-KEM-768 encapsulation and ML-DSA-65 authentication (NIST FIPS 203/204) with AES-256-GCM session keys via HKDF, implemented with Open Quantum Safe liboqs on a Raspberry Pi 4B gateway. Over 500 runs, the post-quantum handshake (Steps 1--6) completes in 2.48 ms, 0.38 ms slower than classical baseline, with PQC computation around 8\% of handshake time at 20 ms simulated round-trip network latency. The gateway sustains 443 sessions/second, 100\% success under 32 concurrent connections, extrapolating to 3,546 sessions/second on a 32-core cloud instance. Five side-channel tests, including verified in-place session-key zeroization and a fixed-vs-random TVLA timing analysis, found no exploitable timing leakage or susceptibility to man-in-the-middle attacks. The architecture is vendor-agnostic and becomes unnecessary once vendors adopt NIST PQC natively.}

\keyword{post-quantum cryptography; ML-KEM; ML-DSA; HVAC security; IoT gateway; PQC migration; smart home; liboqs; FIPS 203; FIPS 204}

\begin{document}

\section{Introduction}
\label{sec:intro}

Smart HVAC systems are now standard in residential and commercial buildings, with hundreds of millions of units deployed globally~\cite{Yuan2023,Panwar2019}. Most modern systems follow a cloud-mediated architecture: a mobile application communicates with the HVAC controller through a vendor-operated cloud service, with TLS protecting traffic in transit~\cite{Panwar2019,Rescorla2018}. The device itself never communicates directly with the user application. This design makes the vendor cloud the natural cryptographic boundary, and it means the security of the entire communication channel depends on whatever asymmetric cryptography TLS uses at that time.

The problem is that HVAC equipment lasts a long time. Legacy controllers are routinely still in service a decade or more after manufacture~\cite{HernandezRamos2020}. Devices purchased today will be operational well into the 2030s and 2040s. This matters because Shor's algorithm, running on a cryptographically relevant quantum computer (CRQC), breaks RSA and ECDH in polynomial time~\cite{Shor1997}. NIST has set a two-stage compliance timeline for quantum-vulnerable algorithms: deprecation for new systems by 2030 and full disallowance by 2035~\cite{NISTIR8547}. The EU Cyber Resilience Act (2024) requires IoT products sold in Europe to support security update mechanisms throughout their lifecycle~\cite{CRA2024}. Even before a CRQC exists, the Harvest Now, Decrypt Later (HNDL) threat is real: an adversary can capture and store encrypted HVAC traffic today and decrypt it retroactively once quantum capability is available~\cite{Liu2024}. For devices with 10 to 15 year lifespans, that window is already open.

The obvious solution would be to update the HVAC controller firmware with post-quantum cryptography. This is not feasible in practice. Three separate constraints prevent it. First, legacy HVAC firmware is signed with proprietary keys and cannot be updated by third parties, and many deployed units lack reliable over-the-air update infrastructure~\cite{HernandezRamos2020,Catuogno2023}. Second, HVAC microcontrollers do not have enough RAM for any ML-KEM variant. We measured this directly: an ESP32-S3, the most capable MCU class used in modern HVAC systems, has 339 KB of free heap at runtime, well beyond RFC 7228's Class 2 constrained-node reference point (~50 KiB RAM)~\cite{RFC7228}, meaning this is not a marginal or unusually constrained device by IoT standards. ML-KEM-768 requires 900 KB for key generation, and malloc(900 KB) fails. Third, even if memory were not a constraint, the compute overhead for asymmetric operations on these devices is already unacceptable. ECDH-P256 takes 111.93 ms in software on the ESP32-S3, compared to 0.032 ms for hardware-accelerated AES-128. Asymmetric cryptography of any kind needs to be offloaded from the device.

Several related works address nearby problems but not this one. \citet{Malina2024} propose quantum-resistant MQTT security but require changes to both the client and broker. \citet{Varanasi2025} evaluate six NIST PQC KEMs integrated into MQTT across thousands of configurations, but again assume that endpoint software can be modified. \citet{Chhetri2026} benchmarks ML-KEM and ML-DSA on an ARM Cortex-M0+ (RP2040, 264 KB SRAM) and reports that ML-KEM-512 completes a full key exchange in 35.7 ms. Our ESP32-S3 measurements extend this line of work to the specific hardware class used in HVAC. \citet{Fitzgibbon2024} benchmark liboqs on Raspberry Pi 4B and confirm gateway suitability, but do not address how to connect a PQC-capable gateway to a legacy device ecosystem that cannot be modified. No existing work combines non-invasive integration, the consumer HVAC domain, vendor independence, and empirical validation of a working system.

We propose a PQC proxy layer interposed between the mobile application and the vendor cloud. The proxy performs ML-KEM-768 key encapsulation and ML-DSA-65 authentication per NIST FIPS 203 and FIPS 204~\cite{FIPS203,FIPS204}, and derives AES-256-GCM session keys via HKDF~\cite{RFC5869}. The HVAC device, vendor cloud, and all existing infrastructure are unchanged. We implement this using the Open Quantum Safe liboqs library~\cite{liboqs2024,OQS_Python2024} on a Raspberry Pi 4B gateway and evaluate it with real hardware measurements.

The specific contributions of this paper are:

\begin{enumerate}
\item \textbf{Architecture.} A four-tier system architecture for non-invasive PQC integration into the cloud-mediated HVAC communication model, with no changes required to the device, its firmware, or the vendor cloud (Section~\ref{sec:architecture}).
\item \textbf{Implementation.} A working prototype of the proposed proxy using the Open Quantum Safe liboqs library on Raspberry Pi 4B, providing a deployable reference for the home gateway deployment model (Section~\ref{sec:implementation}).
\item \textbf{Empirical benchmarking.} Measurements of ML-KEM-512/768/1024 and ML-DSA-44/65/87 against an ECDH-X25519 and Ed25519 baseline across five operations over 500 runs each, showing a Steps 1--6 handshake overhead of 0.38 ms and network RTT dominance under all realistic conditions (Section~\ref{sec:results}).
\item \textbf{Device-side infeasibility proof.} Direct measurement on an ESP32-S3 confirming that no ML-KEM variant fits in the device's available SRAM, and that ECDH-P256 is already too slow (111.93 ms) on this hardware (Section~\ref{sec:results:esp32}).
\item \textbf{Scalability evaluation.} 443 sessions per second on a four-core Raspberry Pi 4B gateway, bounded memory growth (60 KB over 500 sequential sessions, below the leak-detection threshold), and 100\% session success under 32 concurrent connections, extrapolating to 3,546 sessions per second on a 32-core cloud instance (Section~\ref{sec:results:throughput}).
\item \textbf{Side-channel analysis.} Five empirical tests: memory forensics verifying in-place session-key zeroization, MITM simulation confirming ML-DSA-65 blocks impersonation, nonce collision testing, timing-consistency measurement relative to an AES-256-GCM baseline, and a fixed-vs-random TVLA timing leakage assessment~\cite{Goodwill2011} on the architecture's decapsulation path (Section~\ref{sec:security:sidechannel}).
\end{enumerate}

The rest of the paper is structured as follows. Section~\ref{sec:related} covers related work. Section~\ref{sec:background} provides background on the quantum threat, NIST PQC standards, the HVAC communication model, and the implementation platform. Section~\ref{sec:architecture} describes the system architecture. Section~\ref{sec:implementation} covers the implementation. Section~\ref{sec:setup} details the experimental setup. Section~\ref{sec:results} presents results. Section~\ref{sec:security} covers security analysis. Section~\ref{sec:discussion} discusses the findings. Section~\ref{sec:limitations} covers limitations and future work. Section~\ref{sec:conclusion} concludes.

\section{Related Work}
\label{sec:related}
 
\subsection{PQC Performance on Constrained and Gateway Hardware}
\label{sec:related:constrained}
 
A growing body of work benchmarks post-quantum algorithms on hardware relevant to IoT deployment. Fitzgibbon and Ottaviani~\cite{Fitzgibbon2024} measure liboqs on a Raspberry Pi 4B and confirm that ML-KEM and ML-DSA complete key operations in the sub-millisecond range on ARM Cortex-A72 hardware, establishing that gateway-class devices can run PQC without prohibitive overhead. Lopez et al.~\cite{Lopez2025} implement CRYSTALS-Kyber, BIKE, and HQC on Raspberry Pi hardware and demonstrate that PQC integration on constrained IoT platforms is practical in terms of computational overhead, memory usage, and energy consumption. Chhetri~\cite{Chhetri2026} benchmarks ML-KEM and ML-DSA on an ARM Cortex-M0+ (RP2040) with 264 KB of SRAM across all three security levels, reporting a full ML-KEM-512 key exchange completing in 35.7 ms. Tasopoulos et al.~\cite{Tasopoulos2023} evaluate post-quantum TLS 1.3 energy consumption on resource-constrained embedded devices, finding that PQC-integrated TLS can be comparable to, and in some configurations more energy-efficient than, classical TLS 1.3, depending on device role and authentication mode. Segatz and Al Hafiz~\cite{Segatz2025} implement CRYSTALS-Kyber on an ESP32 using hand-partitioned dual-core execution and hardware coprocessor offload, achieving up to a 1.84$\times$ Kyber encapsulation speedup over a single-core baseline implementation. The consistent finding across this body of work is that gateway-class hardware such as Raspberry Pi can run PQC comfortably, while microcontroller-class hardware faces fundamental memory barriers that become worse as security level increases.
 
\subsection{Quantum-Resistant IoT Communication Protocols}
\label{sec:related:protocols}
 
Several works specifically address post-quantum security for IoT communication protocols. Malina et al.~\cite{Malina2024} propose a quantum-resistant security scheme for MQTT that supports secure broadcast for IoT services. Their scheme provides meaningful security guarantees but requires modification of both the client and broker endpoints, which is infeasible for HVAC deployments where neither the controller firmware nor the vendor cloud can be modified. Varanasi~\cite{Varanasi2025} presents a comprehensive performance evaluation of six NIST-standardised and candidate PQC KEMs integrated directly into MQTT across 4,608 experimental configurations spanning diverse network conditions. This is the closest existing study to our network latency analysis, but again assumes that MQTT endpoints can be updated. Liu et al.~\cite{Liu2024} survey PQC performance optimisation for IoT broadly, covering lattice-based, code-based, and hash-based schemes, and conclude that ML-KEM represents the most practical choice for constrained IoT deployments given its balance of performance and key sizes.
 
\subsection{PQC for Industrial and Medical IoT}
\label{sec:related:iiot}
 
In domains beyond smart home, related work has examined PQC integration for industrial and medical IoT. Shahid et al.~\cite{Shahid2026} integrate ML-KEM and ML-DSA into TLS 1.3 for industrial IoT authentication on gateway-class hardware, demonstrating feasibility at the TLS layer, though their work requires modification of TLS endpoints and gateway infrastructure rather than a transparent proxy in front of an unchanged vendor cloud. Nejatollahi et al.~\cite{Nejatollahi2019} survey lattice-based cryptography implementations across a range of hardware platforms and identify that memory and side-channel resistance are the two dominant constraints for practical deployment. Amiriara et al.~\cite{Amiriara2025} address PQC security in edge computing scenarios for resource-constrained devices, proposing offloading architectures that share a conceptual similarity with our proxy approach but do not target the HVAC domain or the specific cloud-mediated architectural constraint we address. Basile et al.~\cite{Basile2024} study security at the edge for resource-limited IoT devices and recommend gateway-mediated security as the most practical path for devices that cannot be updated, which directly supports the architectural premise of our work.
 
\subsection{Smart Home Security and HVAC Architectures}
\label{sec:related:smarthome}
 
Yuan et al.~\cite{Yuan2023} provide a comprehensive survey of smart home security across platform, device, and communication layers, identifying vendor heterogeneity and unclear responsibility boundaries as recurring root causes of insecurity. Panwar et al.~\cite{Panwar2019} survey smart home privacy and security in cloud-assisted home area networks, documenting reliance on classical TLS~\cite{Rescorla2018} for remote access to connected devices. Lin and Bergmann~\cite{Lin2016} highlight fixed firmware and limited update services as persistent liabilities in smart home IoT and argue for gateway-centric architectures with automated firmware maintenance. Fernandez-Carames~\cite{FernandezCarames2020} surveys quantum-resistant cryptosystems for IoT and discusses edge and gateway tiers as deployment options for offloading heavy cryptography from constrained devices, though without a specific HVAC implementation.
 
\subsection{PQC Migration Strategies and Standardisation}
\label{sec:related:migration}
 
NIST finalised ML-KEM (FIPS 203)~\cite{FIPS203} and ML-DSA (FIPS 204)~\cite{FIPS204} as the primary post-quantum standards in August 2024, with NIST IR 8547~\cite{NISTIR8547} outlining a transition timeline requiring federal migration by 2035. The EU Cyber Resilience Act~\cite{CRA2024}, which entered into force in December 2024 with full compliance required by 2027, mandates that IoT devices sold in Europe support security update mechanisms throughout their lifecycle. Paquin et al.~\cite{Paquin2020} benchmark PQC in TLS and find that handshake latency overhead is dominated by larger key and signature sizes rather than computation time at the gateway, consistent with our network latency findings. Stebila and Mosca~\cite{Stebila2017} establish the theoretical and practical foundations of the Open Quantum Safe project, which provides the liboqs library used in our implementation. The algebraic structure underlying ML-KEM security is analysed by Peikert and Pepin~\cite{Peikert2019}, who examine the structured LWE assumptions on which CRYSTALS-Kyber, the predecessor to ML-KEM, is built.
 
\subsection{Research Gap}
 
Table~\ref{tab:related} summarises the positioning of our work relative to the literature. The key distinction is the combination of constraints that our work simultaneously satisfies: the HVAC device is unmodified, the vendor cloud is unmodified, no vendor cooperation is required, and the system is empirically evaluated on real hardware including a working implementation and a device-side infeasibility proof. No existing work satisfies all of these simultaneously.
 
\begin{table*}[h]
\centering

\resizebox{\textwidth}{!}{%
\renewcommand{\arraystretch}{1.2}
\begin{tabular}{lcccccc}
\toprule
\textbf{Work} & \textbf{No Device} & \textbf{No Cloud} & \textbf{Vendor} & \textbf{HVAC} & \textbf{PQC} & \textbf{Empirical} \\
& \textbf{Change} & \textbf{Change} & \textbf{Agnostic} & \textbf{Domain} & \textbf{Standard} & \textbf{Validation} \\
\midrule
Malina et al.~\cite{Malina2024} & $\times$ & $\times$ & $\sim$ & $\times$ & \checkmark & $\sim$ \\
Varanasi~\cite{Varanasi2025} & $\times$ & $\times$ & \checkmark & $\times$ & \checkmark & \checkmark \\
Chhetri~\cite{Chhetri2026} & \checkmark & \checkmark & \checkmark & $\times$ & \checkmark & \checkmark \\
Fitzgibbon \& Ottaviani~\cite{Fitzgibbon2024} & \checkmark & \checkmark & \checkmark & $\times$ & \checkmark & \checkmark \\
Shahid et al.~\cite{Shahid2026} & $\times$ & $\times$ & $\sim$ & $\times$ & \checkmark & \checkmark \\
Basile et al.~\cite{Basile2024} & \checkmark & \checkmark & \checkmark & $\times$ & $\sim$ & $\sim$ \\
Fernandez-Carames~\cite{FernandezCarames2020} & $\sim$ & \checkmark & \checkmark & $\times$ & $\sim$ & $\times$ \\
\midrule
\textbf{This work} & \checkmark & \checkmark & \checkmark & \checkmark & \checkmark & \checkmark \\
\bottomrule
\end{tabular}%
}

\vspace{6pt}
\captionsetup{justification=centering}
\caption{Comparison of related work. Symbols: \checkmark = addressed, $\times$ = not addressed, $\sim$ = partially addressed.}
\label{tab:related}
\end{table*}

\section{Background}
\label{sec:background}

\subsection{Quantum Threat to Asymmetric Cryptography}
\label{sec:background:threat}

The security of classical asymmetric cryptography rests on computational hardness assumptions that hold against classical computers but not against quantum ones. RSA and Diffie-Hellman variants, including ECDH, derive their security from the difficulty of integer factorisation and discrete logarithm problems respectively. Shor's algorithm solves both problems in polynomial time on a quantum computer~\cite{Shor1997}, which means any system relying on these primitives for key exchange or digital signatures becomes insecure once a sufficiently powerful quantum computer exists. Grover's algorithm~\cite{Grover1996} provides a quadratic speedup for unstructured search, which halves the effective security of symmetric keys and hash functions, but this is addressed by doubling key lengths rather than replacing the algorithm. The critical concern for long-lived systems is not the current state of quantum hardware but the trajectory. NIST IR 8547 sets a 2030 deprecation deadline and a 2035 disallowance deadline for RSA, ECDH, and related quantum-vulnerable algorithms in federal systems~\cite{NISTIR8547}, and adversaries operating HNDL strategies are already collecting encrypted data today in anticipation of a future cryptographically relevant quantum computer.

\subsection{NIST Post-Quantum Cryptography Standards}
\label{sec:background:nist}

Following a multi-year evaluation process, NIST published three post-quantum cryptography standards in August 2024. FIPS 203~\cite{FIPS203} specifies ML-KEM (Module Lattice-based Key Encapsulation Mechanism), a key encapsulation mechanism derived from CRYSTALS-Kyber. ML-KEM provides three security levels: ML-KEM-512 (Level 1, roughly equivalent to AES-128), ML-KEM-768 (Level 3, equivalent to AES-192), and ML-KEM-1024 (Level 5, equivalent to AES-256). Security is based on the Module Learning With Errors (MLWE) problem~\cite{Peikert2019}, which is believed to be hard for both classical and quantum computers. FIPS 204~\cite{FIPS204} specifies ML-DSA (Module Lattice-based Digital Signature Algorithm), a digital signature scheme derived from CRYSTALS-Dilithium. ML-DSA provides three parameter sets: ML-DSA-44, ML-DSA-65, and ML-DSA-87, corresponding to the same three security levels. Security is based on the hardness of MLWE and Module Short Integer Solution (MSIS) problems. FIPS 205~\cite{FIPS205} specifies SLH-DSA, a stateless hash-based signature scheme that provides a security guarantee independent of lattice assumptions. This work uses ML-KEM-768 and ML-DSA-65, providing NIST Level 3 security throughout, as this offers a reasonable balance between security margin and performance overhead for gateway deployment.

\subsection{Smart Home HVAC Communication Model}
\label{sec:background:hvac}

Contemporary smart HVAC systems follow a four-tier cloud-mediated communication architecture consistent with cloud-assisted smart home models surveyed by Panwar et al.~\cite{Panwar2019} and Yuan et al.~\cite{Yuan2023}. The four tiers are: the user mobile application, the vendor cloud service, the home network gateway or router, and the HVAC controller device. The mobile application communicates exclusively with the vendor cloud over TLS~\cite{Rescorla2018}. The vendor cloud translates user commands and relays them to the HVAC controller, typically over a proprietary protocol or a simplified IoT messaging protocol such as MQTT. The HVAC controller never communicates directly with the user application. This architecture has two important implications for security. First, all asymmetric cryptography occurs at the TLS layer between the mobile application and the vendor cloud. The device itself only needs to authenticate to its paired cloud service using credentials established at installation time. Second, under this model the vendor cloud typically processes user commands and behavioural data server-side before relaying control instructions to the device, making it a primary data custodian even though users interact only with the mobile application. The proxy architecture proposed in this paper targets specifically the link between the mobile application and the vendor cloud, which is where TLS is already in use and where the cryptographic migration can be performed transparently.

\subsection{Implementation Platform and Libraries}
\label{sec:background:platform}

The Open Quantum Safe (OQS) project~\cite{Stebila2017} provides liboqs, an open-source C library implementing NIST-standardised and candidate post-quantum algorithms~\cite{liboqs2024}. liboqs includes constant-time implementations of ML-KEM and ML-DSA derived from the NIST reference implementations, with AArch64-specific optimisations that improve performance on ARM-based hardware. The library is maintained under the Linux Foundation Post-Quantum Cryptography Alliance and is the most widely used open-source PQC library for research and prototyping. We use liboqs version 0.12 via the official Python binding~\cite{OQS_Python2024}, which provides a clean interface for key generation, encapsulation, decapsulation, signing, and verification without requiring direct C interoperability in the application layer.

The gateway hardware platform is a Raspberry Pi 4B with 4 GB of RAM running a 64-bit Raspberry Pi OS (Debian Bookworm). The processor is an ARM Cortex-A72 quad-core at 1800 MHz. This platform represents a realistic home gateway deployment: it is commercially available for approximately \$55 USD, runs a full Linux distribution, and has sufficient RAM and compute for all PQC operations without specialised hardware.

The device-side platform is an ESP32-S3 development board with an Xtensa dual-core LX7 processor at 240 MHz and 380 KB of total SRAM. RFC 7228 defines a Class 2 constrained node as having roughly 50 KiB of RAM and 250 KiB of flash~\cite{RFC7228}, and notes explicitly that it does not attempt to classify devices with capabilities significantly beyond Class 2. The ESP32-S3's SRAM exceeds that reference point by roughly 7$\times$, making it representative of the upper end, not the constrained end, of microcontrollers found in modern commercial HVAC systems. All ESP32-S3 benchmarks in this paper are measured on physical hardware using the Arduino framework with mbedTLS for classical cryptographic operations.

\section{System Architecture}
\label{sec:architecture}

\subsection{Design Principles}
\label{sec:arch:principles}

The architecture is governed by three design principles, each derived directly from the constraints established in Section~\ref{sec:background}. \textbf{Non-invasive integration} requires that no modification be made to the HVAC device, its firmware, or the vendor cloud. This follows from the device-side infeasibility established empirically in Section~\ref{sec:background:hvac} and confirmed on real hardware in Section~\ref{sec:results:esp32}. \textbf{Vendor independence} requires that the architecture function without cooperation from the HVAC manufacturer, since commercial vendors such as Nest, Ecobee, and Honeywell operate closed, proprietary cloud infrastructure that cannot be modified by a third party and have not announced native ML-KEM support. \textbf{Cryptographic agility} requires that the component performing post-quantum operations be independently updatable, so that a future revision to NIST's selected algorithms does not require any change to the HVAC device, the vendor cloud, or the mobile application. These three principles jointly motivate the insertion of a dedicated proxy layer, rather than embedding post-quantum cryptography directly into any existing system component.

\subsection{System Overview}
\label{sec:arch:overview}

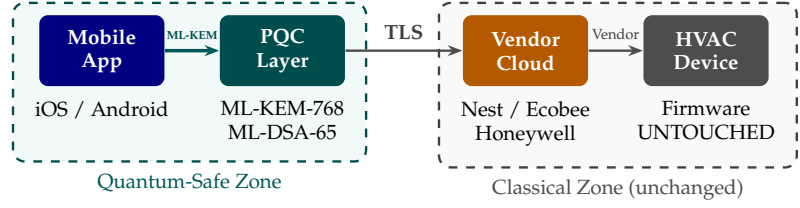
\begin{figure}[t]
\centering
\resizebox{0.75\linewidth}{!}{%
\begin{tikzpicture}[
    node distance = 0.55cm and 0.65cm,
    box/.style = {
        rectangle, rounded corners=3pt,
        minimum width=1.5cm, minimum height=0.85cm,
        align=center, font=\scriptsize\bfseries,
        text=white, inner sep=4pt
    },
    subtext/.style = {
        font=\scriptsize, text=black, align=center
    },
    arr/.style = {
        -{Stealth[length=5pt]}, thick
    },
    zone/.style = {
        rectangle, rounded corners=4pt,
        draw, dashed, thick,
        inner sep=4pt
    }
]
\node[box, fill=blue!50!black] (app) {Mobile\\App};
\node[subtext, below=0.05cm of app] (appsub) {iOS / Android};
\node[box, fill=teal!60!black, right=0.65cm of app] (pqc) {PQC\\Layer};
\node[subtext, below=0.05cm of pqc] (pqcsub) {ML-KEM-768\\ML-DSA-65};
\node[box, fill=orange!70!black, right=1.4cm of pqc] (vendor) {Vendor\\Cloud};
\node[subtext, below=0.05cm of vendor] (vendorsub) {Nest / Ecobee\\Honeywell};
\node[box, fill=gray!60!black, right=0.65cm of vendor] (hvac) {HVAC\\Device};
\node[subtext, below=0.05cm of hvac] (hvacsub) {Firmware\\UNTOUCHED};
\draw[arr, teal!70!black, line width=1.2pt]
    (app.east) -- (pqc.west)
    node[midway, above, font=\fontsize{4}{6}\selectfont\bfseries, text=teal!60!black] {ML-KEM};
\draw[arr, gray!60!black]
    (pqc.east) -- (vendor.west)
    node[midway, above, font=\scriptsize\bfseries, text=gray!50!black] {TLS};
\draw[arr, gray!60!black]
    (vendor.east) -- (hvac.west)
    node[midway, above, font=\tiny, text=gray!50!black] {Vendor};
\begin{scope}[on background layer]
  \node[zone, fill=teal!5, draw=teal!60!black,
        fit=(app)(appsub)(pqc)(pqcsub),
        label={[font=\scriptsize, text=teal!70!black]below:Quantum-Safe Zone}] (qzone) {};
  \node[zone, fill=gray!5, draw=gray!50!black,
        fit=(vendor)(vendorsub)(hvac)(hvacsub),
        label={[font=\scriptsize, text=gray!60!black]below:Classical Zone (unchanged)}] (czone) {};
\end{scope}
\end{tikzpicture}%
}

\vspace{6pt}
\captionsetup{justification=centering}
\caption{Proposed four-tier system architecture. The PQC Layer terminates the quantum-safe session with the Mobile App using ML-KEM-768 and ML-DSA-65. All downstream components remain unmodified.}
\label{fig:architecture}
\end{figure}

The proposed architecture introduces a four-tier communication model, shown in Figure~\ref{fig:architecture}. The tiers are (1) the \textit{Mobile Application}, running on the user's smartphone, (2) the \textit{PQC Layer}, a dedicated proxy component performing post-quantum key establishment and authentication, (3) the \textit{Vendor Cloud}, the existing HVAC manufacturer's cloud infrastructure, unmodified, and (4) the \textit{HVAC Controller}, the physical device on the user's premises, unmodified.

In the existing classical deployment, the mobile application connects directly to the vendor cloud, with the connection secured by TLS using ECDH key exchange and ECDSA authentication, both vulnerable to Shor's algorithm (Section~\ref{sec:background:threat}). The proposed architecture inserts the PQC Layer between the mobile application and the vendor cloud. The mobile application no longer connects to the vendor cloud directly. Instead, it establishes a quantum-safe session with the PQC Layer, which then forwards commands to the vendor cloud over the existing, unmodified classical TLS connection. The quantum-safe zone therefore spans only the link between the mobile application and the PQC Layer. The link between the PQC Layer and the vendor cloud, and between the vendor cloud and the HVAC device, remains classical. This bounded scope is a deliberate design constraint, not an oversight: it is precisely the segment of the communication path that can be secured without vendor cooperation or device access, and it is the segment most exposed to Harvest-Now-Decrypt-Later collection, since it carries the session establishment traffic.

Directing the mobile application's traffic to the PQC Layer rather than the vendor cloud requires only a one-time endpoint configuration change in the mobile application, distributed through the vendor's existing app store channel, or, in a home-gateway deployment, a local DNS override at the router level with no change to the application binary at all.

\subsection{Justification for a Separate PQC Layer}
\label{sec:arch:justification}

An alternative design would integrate post-quantum cryptography directly within the vendor cloud, eliminating the separate proxy. We reject this alternative for two reasons.

First, vendor cloud infrastructure is proprietary and inaccessible. Commercial HVAC platforms operate closed server-side systems, and no commercial vendor has announced ML-KEM integration at the time of writing. An architecture that depends on vendor cooperation is not deployable today, regardless of its technical merits.

Second, a separate PQC Layer provides cryptographic agility and vendor independence that a vendor-embedded solution cannot. When NIST revises or supersedes ML-KEM or ML-DSA, the PQC Layer is the only component requiring modification. No vendor engineering effort or coordinated device update is needed. A single PQC Layer deployment can additionally serve users of multiple HVAC vendors without modification, whereas vendor-specific integration would require an independent implementation per ecosystem.

This design does introduce a new network node, and with it a potential single point of failure. We treat this as an acknowledged deployment consideration rather than a fundamental limitation: production deployments should run the PQC Layer with standard high-availability practices, including load balancing and geographic redundancy, consistent with conventional cloud proxy deployment. Section~\ref{sec:security:limitations} discusses this trade-off, and Section~\ref{sec:results:throughput} presents measured throughput and concurrent-session behaviour of a single-instance deployment as a baseline.

\subsection{Authentication and Key Establishment Flow}
\label{sec:arch:flow}

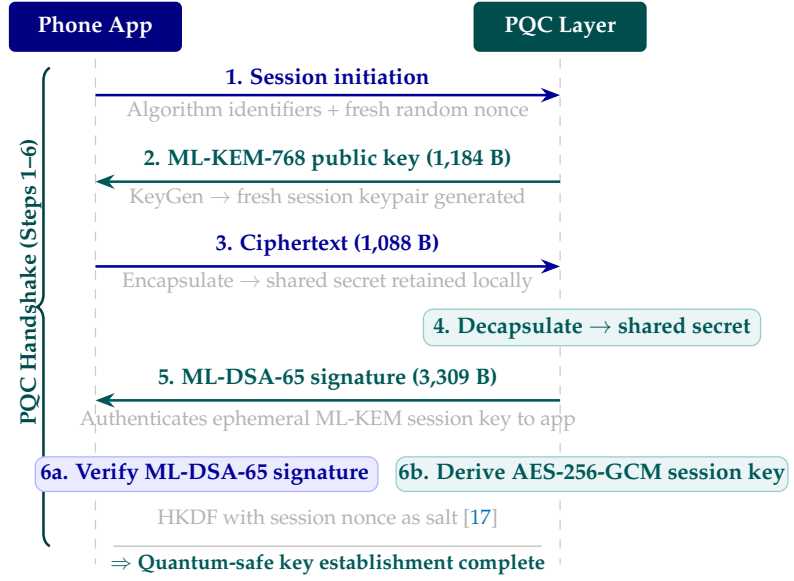
\begin{figure}[t]
\centering
\resizebox{0.75\linewidth}{!}{%
\begin{tikzpicture}[
    font=\scriptsize,
    every node/.style={inner sep=3pt},
    arr/.style={-{Stealth[length=6pt]}, thick},
    box/.style={rectangle, fill=#1, text=white,
                font=\scriptsize\bfseries,
                minimum width=1.9cm, minimum height=0.55cm,
                rounded corners=2pt, align=center}
]
\def\xPhone{-0.1}
\def\xPQC{5}
\node[box=blue!50!black]  at (\xPhone, 0) (Phone) {Phone App};
\node[box=teal!60!black]  at (\xPQC,   0) (PQC)   {PQC Layer};
\draw[gray!40, thin, dashed] (\xPhone,-0.35) -- (\xPhone,-5.9);
\draw[gray!40, thin, dashed] (\xPQC,  -0.35) -- (\xPQC,  -5.9);
\draw[teal!50!black, thick, decorate,
      decoration={brace, amplitude=4pt, mirror}]
    (\xPhone-0.5,-0.45) -- (\xPhone-0.5,-5.7)
    node[very near start, left=6pt, rotate=90,
         font=\scriptsize\bfseries, text=teal!60!black,
         align=center] {PQC Handshake (Steps 1--6)};
\draw[arr, blue!60!black]
    (\xPhone,-0.75) -- (\xPQC,-0.75)
    node[midway, above, text=blue!60!black, font=\scriptsize\bfseries]
         {\textbf{1.} Session initiation};
\node[font=\fontsize{6.5}{7}\selectfont, text=gray!60] at
    ({(\xPhone+\xPQC)/2},-0.93) {Algorithm identifiers + fresh random nonce};
\draw[arr, teal!70!black]
    (\xPQC,-1.70) -- (\xPhone,-1.70)
    node[midway, above, text=teal!70!black, font=\scriptsize\bfseries]
         {\textbf{2.} ML-KEM-768 public key (1,184 B)};
\node[font=\fontsize{6.5}{7}\selectfont, text=gray!60] at
    ({(\xPhone+\xPQC)/2},-1.88) {KeyGen $\rightarrow$ fresh session keypair generated};
\draw[arr, blue!60!black]
    (\xPhone,-2.63) -- (\xPQC,-2.63)
    node[midway, above, text=blue!60!black, font=\fontsize{7}{7}\selectfont\bfseries]
         {\textbf{3.} Ciphertext (1,088 B)};
\node[font=\fontsize{6.5}{7}\selectfont, text=gray!60] at
    ({(\xPhone+\xPQC)/2},-2.81) {Encapsulate $\rightarrow$ shared secret retained locally};
\draw[rounded corners=3pt, fill=teal!8, draw=teal!40]
    (\xPQC-1.5,-3.1) rectangle (\xPQC+2.2,-3.5);
\node[text=teal!70!black, font=\scriptsize\bfseries] at
    (\xPQC+0.35,-3.3) {\textbf{4.} Decapsulate $\rightarrow$ shared secret};
\draw[arr, teal!70!black]
    (\xPQC,-4.10) -- (\xPhone,-4.10)
    node[midway, above, text=teal!70!black, font=\scriptsize\bfseries]
         {\textbf{5.} ML-DSA-65 signature (3,309 B)};
\node[font=\fontsize{6.5}{7}\selectfont, text=gray!60] at
    ({(\xPhone+\xPQC)/2},-4.32) {Authenticates ephemeral ML-KEM session key to app};
\draw[rounded corners=3pt, fill=blue!8, draw=blue!30]
    (\xPhone-0.66,-4.72) rectangle (\xPhone+3.1,-5.10);
\node[text=blue!60!black, font=\scriptsize\bfseries] at
    (\xPhone+1.2,-4.91) {\textbf{6a.} Verify ML-DSA-65 signature};
\draw[rounded corners=3pt, fill=teal!8, draw=teal!40]
    (\xPQC-1.8,-4.72) rectangle (\xPQC+2.5,-5.10);
\node[text=teal!70!black, font=\scriptsize\bfseries] at
    (\xPQC+0.35,-4.91) {\textbf{6b.} Derive AES-256-GCM session key};
\node[font=\fontsize{6.5}{7}\selectfont, text=gray!60, align=center] at
    ({(\xPhone+\xPQC)/2},-5.40) {HKDF with session nonce as salt~\cite{RFC5869}};
\draw[gray!50, thin] (\xPhone+0.2,-5.70) -- (\xPQC-0.2,-5.70);
\node[font=\fontsize{6.5}{7}\selectfont\bfseries, text=teal!70!black] at
    ({(\xPhone+\xPQC)/2},-5.9) {$\Rightarrow$ Quantum-safe key establishment complete};
\end{tikzpicture}%
}
\caption{Post-quantum handshake flow (Steps~1--6). The PQC Layer generates a fresh ML-KEM-768 session keypair per session. The ML-DSA-65 signature authenticates that ephemeral key to the Mobile Application via a pre-installed pinned verification key. Command forwarding is shown in Figure~\ref{fig:commandflow}.}
\label{fig:handshake}
\end{figure}

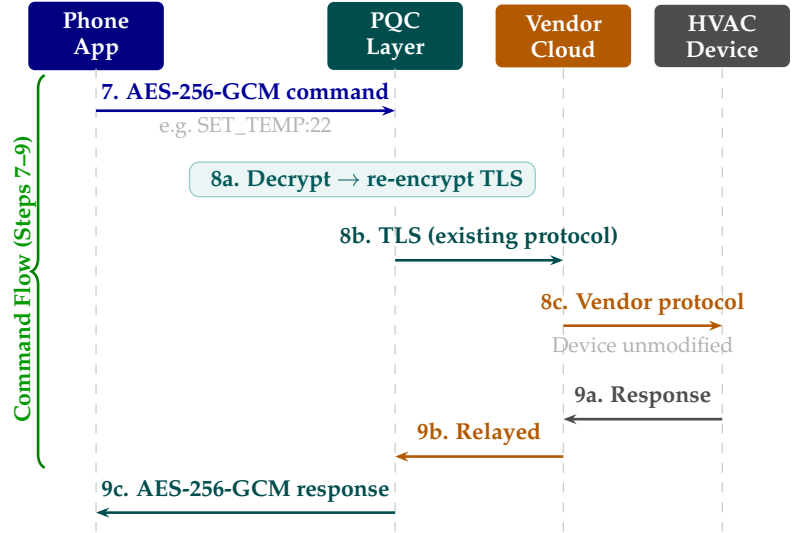
\begin{figure}[t]
\centering
\resizebox{0.75\linewidth}{!}{%
\begin{tikzpicture}[
    font=\scriptsize,
    every node/.style={inner sep=3pt},
    arr/.style={-{Stealth[length=4pt]}, thick},
    box/.style={rectangle, fill=#1, text=white,
                font=\scriptsize\bfseries,
                minimum width=1.45cm, minimum height=0.52cm,
                rounded corners=2pt, align=center}
]
\def\xPhone{0.0}
\def\xPQC{3.2}
\def\xVendor{5}
\def\xDevice{6.7}
\node[box=blue!50!black]   at (\xPhone,  0) {Phone\\App};
\node[box=teal!60!black]   at (\xPQC,   0) {PQC\\Layer};
\node[box=orange!70!black] at (\xVendor, 0) {Vendor\\Cloud};
\node[box=gray!60!black]   at (\xDevice, 0) {HVAC\\Device};
\foreach \x in {\xPhone, \xPQC, \xVendor, \xDevice}{
    \draw[gray!40, thin, dashed] (\x,-0.35) -- (\x,-5.4);}
\draw[green!50!black, thick, decorate,
      decoration={brace, amplitude=4pt, mirror}]
    (-0.55,-0.42) -- (-0.55,-4.62)
    node[very near start, left=6pt, rotate=90,
         font=\scriptsize\bfseries, text=green!60!black,
         align=center] {Command Flow (Steps 7--9)};
\draw[arr, blue!60!black]
    (\xPhone,-0.8) -- (\xPQC,-0.8)
    node[midway, above, text=blue!60!black, font=\scriptsize\bfseries]
         {\textbf{7.} AES-256-GCM command};
\node[font=\fontsize{6.5}{7}\selectfont, text=gray!60] at
    ({(\xPhone+\xPQC)/2},-0.98) {e.g.\ SET\_TEMP:22};
\draw[rounded corners=3pt, fill=teal!8, draw=teal!40]
    (\xPQC-2.2,-1.35) rectangle (\xPQC+1.5,-1.72);
\node[text=teal!70!black, font=\scriptsize\bfseries] at
    (\xPQC-0.3,-1.55) {\textbf{8a.} Decrypt $\rightarrow$ re-encrypt TLS};
\draw[arr, teal!60!black]
    (\xPQC,-2.4) -- (\xVendor,-2.4)
    node[midway, above, text=teal!60!black, font=\scriptsize\bfseries]
         {\textbf{8b.} TLS (existing protocol)};
\draw[arr, orange!70!black]
    (\xVendor,-3.1) -- (\xDevice,-3.1)
    node[midway, above, text=orange!70!black, font=\scriptsize\bfseries]
         {\textbf{8c.} Vendor protocol};
\node[font=\fontsize{6.5}{7}\selectfont, text=gray!60] at
    ({(\xVendor+\xDevice)/2},-3.30) {Device unmodified};
\draw[arr, gray!60!black]
    (\xDevice,-4.10) -- (\xVendor,-4.10)
    node[midway, above, text=gray!60!black, font=\scriptsize\bfseries]
         {\textbf{9a.} Response};
\draw[arr, orange!70!black]
    (\xVendor,-4.5) -- (\xPQC,-4.5)
    node[midway, above, text=orange!70!black, font=\scriptsize\bfseries]
         {\textbf{9b.} Relayed};
\draw[arr, teal!60!black]
    (\xPQC,-5.10) -- (\xPhone,-5.10)
    node[midway, above, text=teal!60!black, font=\scriptsize\bfseries]
         {\textbf{9c.} AES-256-GCM response};
\end{tikzpicture}%
}
\caption{Command forwarding phase (Steps~7--9). The PQC Layer decrypts the AES-256-GCM command, re-encrypts it over the existing classical TLS session, and forwards it to the Vendor Cloud, which relays it to the HVAC Device via the existing vendor protocol. The HVAC Device is unaware of the PQC exchange in Figure~\ref{fig:handshake}.}
\label{fig:commandflow}
\end{figure}

The complete protocol comprises nine steps across two phases: a post-quantum handshake between the mobile application and the PQC Layer (Steps 1--6), and a command-forwarding phase (Steps 7--9). Figure~\ref{fig:handshake} illustrates Steps 1--6, and Figure~\ref{fig:commandflow} illustrates Steps 7--9.

\textbf{Step 1: Session initiation.} The mobile application sends a session initiation message to the PQC Layer, containing supported algorithm identifiers and a fresh random nonce.

\textbf{Step 2: ML-KEM KeyGen.} The PQC Layer executes ML-KEM-768 KeyGen, generating a fresh public-private keypair for this session, and transmits the 1{,}184-byte public key to the mobile application.

\textbf{Step 3: Encapsulation.} The mobile application executes ML-KEM-768 Encapsulate using the received public key, producing a 32-byte shared secret and a 1{,}088-byte ciphertext. The ciphertext is transmitted to the PQC Layer. The shared secret is retained locally and never transmitted.

\textbf{Step 4: Decapsulation.} The PQC Layer executes ML-KEM-768 Decapsulate using its private key and the received ciphertext, recovering the identical shared secret. Neither party has transmitted the shared secret itself at any point.

\textbf{Step 5: ML-DSA signature.} The PQC Layer signs a transcript of the handshake, comprising the Step~1 nonce, the ML-KEM public key, and the ciphertext, using its long-term ML-DSA-65 private signing key, and transmits the resulting 3{,}309-byte signature to the mobile application.

\textbf{Step 6: Verification and key derivation.} The mobile application verifies the ML-DSA-65 signature against the PQC Layer's pre-distributed public verification key, which is embedded in the application binary at installation time and pinned against modification. A valid signature proves that the entity that generated the ephemeral ML-KEM keypair in Step~2 is the legitimate PQC Layer, preventing man-in-the-middle substitution of the session public key. The pinned key authenticates each freshly generated session key rather than substituting for it, which is necessary because forward secrecy requires a new ML-KEM keypair every session. Both parties then derive an AES-256-GCM session key from the shared secret using HKDF~\cite{RFC5869}, with the session nonce as salt.

\textbf{Steps 7--9: Command forwarding.} The mobile application encrypts a command using AES-256-GCM with a fresh 96-bit nonce and transmits it to the PQC Layer. The PQC Layer decrypts the command, verifies the authentication tag, checks the nonce against a per-session replay registry, re-encrypts the command over the existing classical TLS connection, and forwards it to the vendor cloud, which relays it to the HVAC device using the existing, unmodified vendor protocol. The response follows the same path in reverse. The HVAC device is unaware at any point that the preceding exchange used post-quantum cryptography. Figure~\ref{fig:commandflow} shows the complete forwarding sequence.

The per-session nonce registry is initialised at the start of each session and cleared on clean termination. If the PQC Layer terminates uncleanly, for example following a crash, the registry is treated as invalidated and the mobile application must re-initiate the handshake from Step~1, using a fresh keypair and nonce space. This prevents replay of messages from a crashed session into a subsequently recovered one. Section~\ref{sec:results:handshake} reports measured latency for Steps~1--6 on real hardware, and Section~\ref{sec:disc:performance} compares this measurement against the literature-derived estimate reported in our earlier conference-stage analysis of this architecture.

\section{Implementation}
\label{sec:implementation}

\subsection{Implementation Stack}
\label{sec:impl:stack}

The PQC Layer is implemented in Python 3.13.5, using the Open Quantum Safe project's \texttt{liboqs} library through its Python wrapper for ML-KEM-768 and ML-DSA-65 operations~\cite{liboqs2024,OQS_Python2024,Stebila2017}. AES-256-GCM encryption, HKDF key derivation, and the classical ECDH and Ed25519 baseline operations use the Python \texttt{cryptography} library's hazmat primitives. All Raspberry Pi 4B benchmarks in this paper run on this stack, on a 64-bit Linux kernel (aarch64), at a fixed CPU frequency of 1800 MHz. The ESP32-S3 device-side measurements use the Arduino framework with mbedTLS for classical cryptographic operations, described further in Section~\ref{sec:impl:esp32}. Algorithm selection, HKDF context strings, and nonce sizes are centralised in a single configuration module, so that every benchmark script in this study draws from the same parameters.

\subsection{PQC Proxy Implementation}
\label{sec:impl:proxy}

The \texttt{PQCProxy} class implements the architecture described in Section~\ref{sec:arch:flow}. A long-term ML-DSA-65 signing keypair is generated once when the proxy is instantiated and persists across sessions. Each session begins with a fresh ML-KEM-768 keypair generated in \texttt{run\_handshake()}, and the corresponding private key is discarded when the session ends, providing the per-session forward secrecy described in Section~\ref{sec:arch:overview}. The \texttt{decapsulate()} method performs decapsulation, ML-DSA-65 signing of the handshake transcript, and HKDF-based session key derivation in a single call, which is what our end-to-end and side-channel timing measurements in Sections~\ref{sec:results} and~\ref{sec:security:sidechannel} refer to as the architecture's decapsulation path, as distinct from the underlying \texttt{liboqs} primitive in isolation.

The derived session key is stored as a mutable byte array rather than an immutable byte string. This is a deliberate implementation choice: Python's built-in \texttt{bytes} type cannot be overwritten in place, so a session key stored that way can never be reliably cleared from memory on session end, only dereferenced and left to the garbage collector's discretion. Storing the key as a mutable buffer allows \texttt{end\_session()} to overwrite every byte with zero before the session state is reset. Section~\ref{sec:security:sidechannel} verifies empirically that this succeeds. Command encryption and decryption use AES-256-GCM with a fresh 96-bit nonce per message, and a per-session nonce registry rejects any previously observed nonce, providing the replay resistance described in Section~\ref{sec:arch:flow}.

\subsection{Classical Baseline Implementation}
\label{sec:impl:baseline}

The classical ECDH baseline mirrors the PQC proxy's structure exactly, so that the two can be compared on equal terms rather than against a differently shaped implementation. It uses X25519 for key exchange in place of ML-KEM-768, Ed25519 for signing in place of ML-DSA-65, and the same HKDF derivation and AES-256-GCM command handling as the PQC proxy. Every operation that is timed on the PQC side in Section~\ref{sec:results} has a directly corresponding operation timed on the classical side, using the same measurement code path.

\subsection{Thread Safety and Session Isolation}
\label{sec:impl:threads}

The concurrent-session benchmarks in Section~\ref{sec:results:throughput} construct one proxy instance per thread, each sharing the same long-term ML-DSA-65 keypair but holding independent per-session state, meaning session keys and nonce registries. The long-term signing key is read-only once generated and is safe to share across threads without locking. Per-session state is never shared between threads, which avoids the race conditions that would otherwise arise from concurrent sessions writing to a shared nonce registry or session key.

\subsection{ESP32-S3 Device Simulation}
\label{sec:impl:esp32}

The ESP32-S3 measurements in Section~\ref{sec:results:esp32} establish device-side infeasibility through two separate checks rather than a full \texttt{liboqs} execution on the device. First, a direct \texttt{malloc(900\,KB)} call tests whether the device can allocate the working memory ML-KEM-768 key generation requires, against the free heap available at runtime. Second, classical operations that the device firmware would otherwise need to perform, AES-128 encryption and ECDH-P256 key generation, are benchmarked directly using mbedTLS, to confirm that the device is capable of ordinary classical cryptography and that the ML-KEM-768 result reflects a memory constraint specific to post-quantum algorithms rather than a general inability to perform cryptographic operations.

\section{Experimental Setup}
\label{sec:setup}

\subsection{Hardware Configuration}
\label{sec:setup:hardware}

The PQC Layer runs on a Raspberry Pi 4B, representing the home-gateway deployment model described in Section~\ref{sec:arch:overview}. The board runs a 64-bit Linux kernel on the aarch64 architecture, with the CPU frequency locked at 1800 MHz for the duration of every benchmark, so that results are not affected by dynamic frequency scaling between runs. Device-side measurements use an ESP32-S3 development board, described in Section~\ref{sec:impl:esp32}, to represent the HVAC controller's compute and memory envelope. A development workstation drives the mobile application simulation used in the end-to-end and network latency benchmarks in Section~\ref{sec:results}.

\subsection{Benchmark Methodology}
\label{sec:setup:methodology}

Every benchmark follows the same measurement procedure. Each configuration is run 50 times as a warm-up, discarded before measurement begins, followed by 500 measured runs, with a 10 millisecond pause between individual runs. Before each benchmark begins, the Raspberry Pi's temperature is checked against a 72\textdegree{}C threshold, and the run is aborted if this threshold is exceeded, to prevent thermal throttling from silently biasing latency measurements. For each measured operation we report the mean, standard deviation, minimum, maximum, and 95th percentile latency in milliseconds, computed from raw nanosecond-resolution timestamps.

\subsection{Energy Estimation Model}
\label{sec:setup:energy}

Direct hardware energy measurement was not available for this study. Energy per operation is instead estimated as the product of measured execution time and a fixed power draw of 0.6 W, derived by linearly scaling the ARM Cortex-A72's published 0.5 W reference power at 1500 MHz~\cite{ARM_CortexA72} to the Raspberry Pi 4B's benchmark operating frequency of 1800 MHz~\cite{RPiFoundation2019}. This is a constant-power model. It captures relative energy cost between operations of different duration, but it cannot detect any difference in power draw between operations that take the same amount of time, since energy is computed as a fixed multiple of time in every case. We report these figures as estimates throughout this paper, and identify direct hardware energy instrumentation as future work in Section~\ref{sec:limitations}.

\subsection{Experimental Configurations}
\label{sec:setup:configs}

We benchmark all three NIST-standardised ML-KEM and ML-DSA security levels, ML-KEM-512 with ML-DSA-44, ML-KEM-768 with ML-DSA-65, and ML-KEM-1024 with ML-DSA-87, against a classical ECDH-X25519 and Ed25519 baseline implemented as described in Section~\ref{sec:impl:baseline}. ML-KEM-768 with ML-DSA-65 is the configuration used throughout the end-to-end, network, stress, and side-channel evaluations, as it is the configuration proposed for the architecture in Section~\ref{sec:architecture}. Network latency is evaluated by simulating round-trip times of 0, 20, 50, 100, and 250 milliseconds, covering the range from an idealised local connection to a degraded broadband or cellular link. Concurrent session handling is evaluated at 1, 2, 4, 8, 16, and 32 simultaneous threads. Side-channel evaluation sample sizes are reported individually alongside each result in Section~\ref{sec:security:sidechannel}, since different tests require different sample sizes to achieve adequate statistical power.

\section{Evaluation Results}
\label{sec:results}

\subsection{Individual Operation Latency}
\label{sec:results:operations}

Table~\ref{tab:latency} reports mean latency, standard deviation, 95th percentile, and estimated energy for every keygen, encapsulation, decapsulation, signing, and verification operation across all four configurations, measured over 500 runs each. ML-KEM operations across all three security levels complete in under 0.55 ms on average, and decapsulation is consistently the fastest ML-KEM operation. Signing is the single most expensive operation in every ML-DSA configuration, exceeding the corresponding KEM operations by a factor of roughly two, and is the dominant contributor to total handshake time established in Section~\ref{sec:results:handshake}.

\begin{table*}[t]
\centering
\renewcommand{\arraystretch}{1.2}
\small
\begin{tabular*}{\textwidth}{@{\extracolsep{\fill}}lccccc@{}}
\toprule
\textbf{Config} & \textbf{Op} & \textbf{Mean (ms)} & \textbf{Std} & \textbf{P95} & \textbf{Energy (mJ)} \\
\midrule
ECDH-X25519 $+$ Ed25519 & keygen & 0.3300 & 0.0865 & 0.5618 & 0.198015 \\
 & exchange & 0.5950 & 0.0929 & 0.8219 & 0.356998 \\
 & sign & 0.2835 & 0.0679 & 0.4620 & 0.170113 \\
 & verify & 0.4434 & 0.0735 & 0.6371 & 0.26603 \\
\midrule
ML-KEM-512 $+$ ML-DSA-44 & keygen & 0.4631 & 0.1553 & 0.8474 & 0.277844 \\
 & encap & 0.4535 & 0.1423 & 0.8202 & 0.272124 \\
 & decap & 0.3439 & 0.1016 & 0.5677 & 0.206337 \\
 & sign & 0.9514 & 0.3274 & 1.6232 & 0.570867 \\
 & verify & 0.6459 & 0.1742 & 1.0841 & 0.387514 \\
\midrule
ML-KEM-768 $+$ ML-DSA-65 & keygen & 0.5292 & 0.2067 & 1.0676 & 0.317532 \\
 & encap & 0.4812 & 0.1339 & 0.7783 & 0.288708 \\
 & decap & 0.3716 & 0.0973 & 0.5963 & 0.22296 \\
 & sign & 1.2209 & 0.4252 & 2.0718 & 0.732533 \\
 & verify & 0.7575 & 0.1780 & 1.1660 & 0.454518 \\
\midrule
ML-KEM-1024 $+$ ML-DSA-87 & keygen & 0.5471 & 0.1826 & 1.0203 & 0.328275 \\
 & encap & 0.4704 & 0.1534 & 0.7862 & 0.282215 \\
 & decap & 0.4301 & 0.1228 & 0.7916 & 0.258054 \\
 & sign & 1.4802 & 0.4760 & 2.3613 & 0.888105 \\
 & verify & 0.9586 & 0.1989 & 1.4738 & 0.575136 \\
\bottomrule
\end{tabular*}

\vspace{6pt}
\captionsetup{justification=centering}
\caption{Mean operation latency (ms) and estimated energy (mJ) on Raspberry Pi~4B. Values: mean~$\pm$~std (500 runs, 50 warm-up, CPU at 1800~MHz).}
\label{tab:latency}
\end{table*}

\begin{figure*}[t]
\centering
\includegraphics[width=1.0\textwidth]{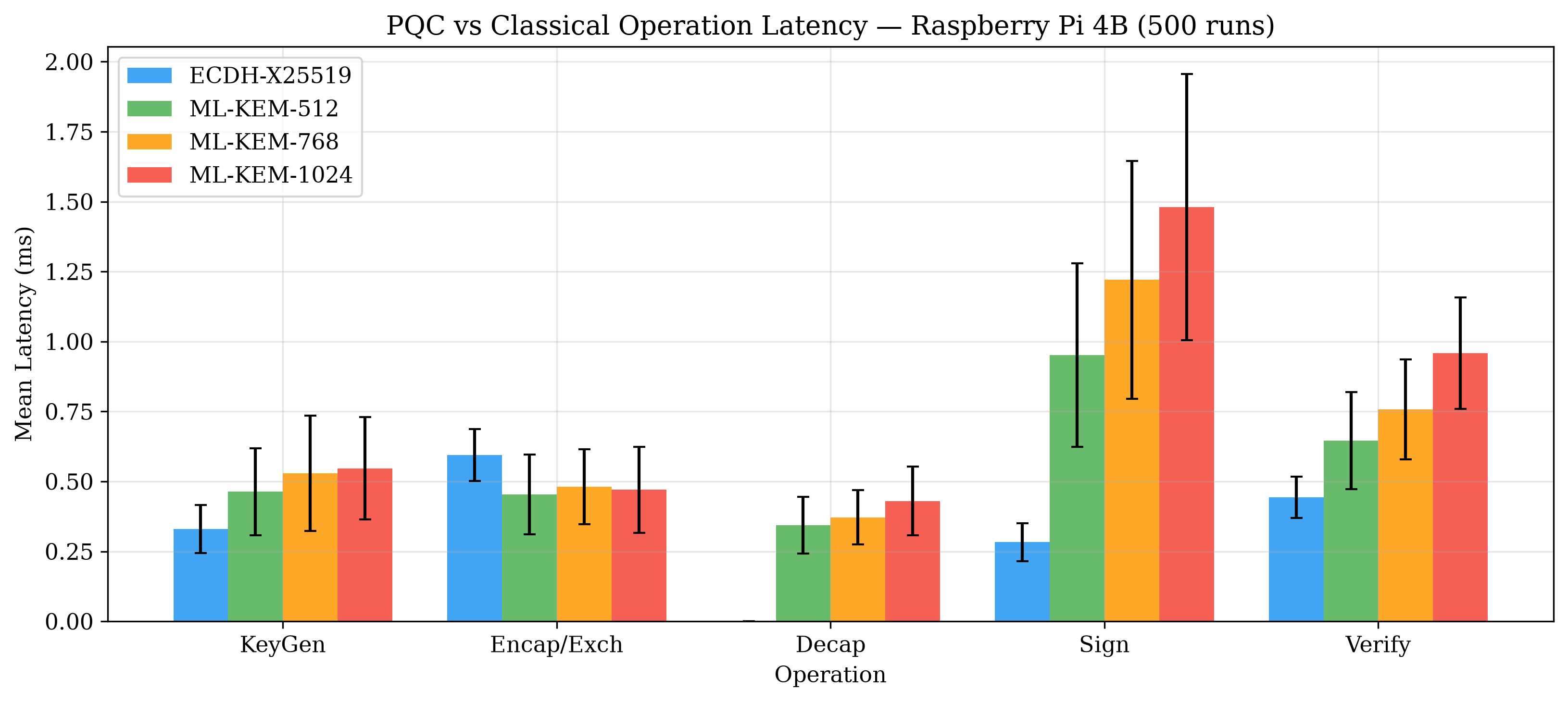}
\captionsetup{justification=centering}
\caption{Mean operation latency by configuration, corresponding to Table~\ref{tab:latency}.}
\label{fig:latency}
\end{figure*}

\begin{figure*}[t]
\centering
\includegraphics[width=1.0\textwidth]{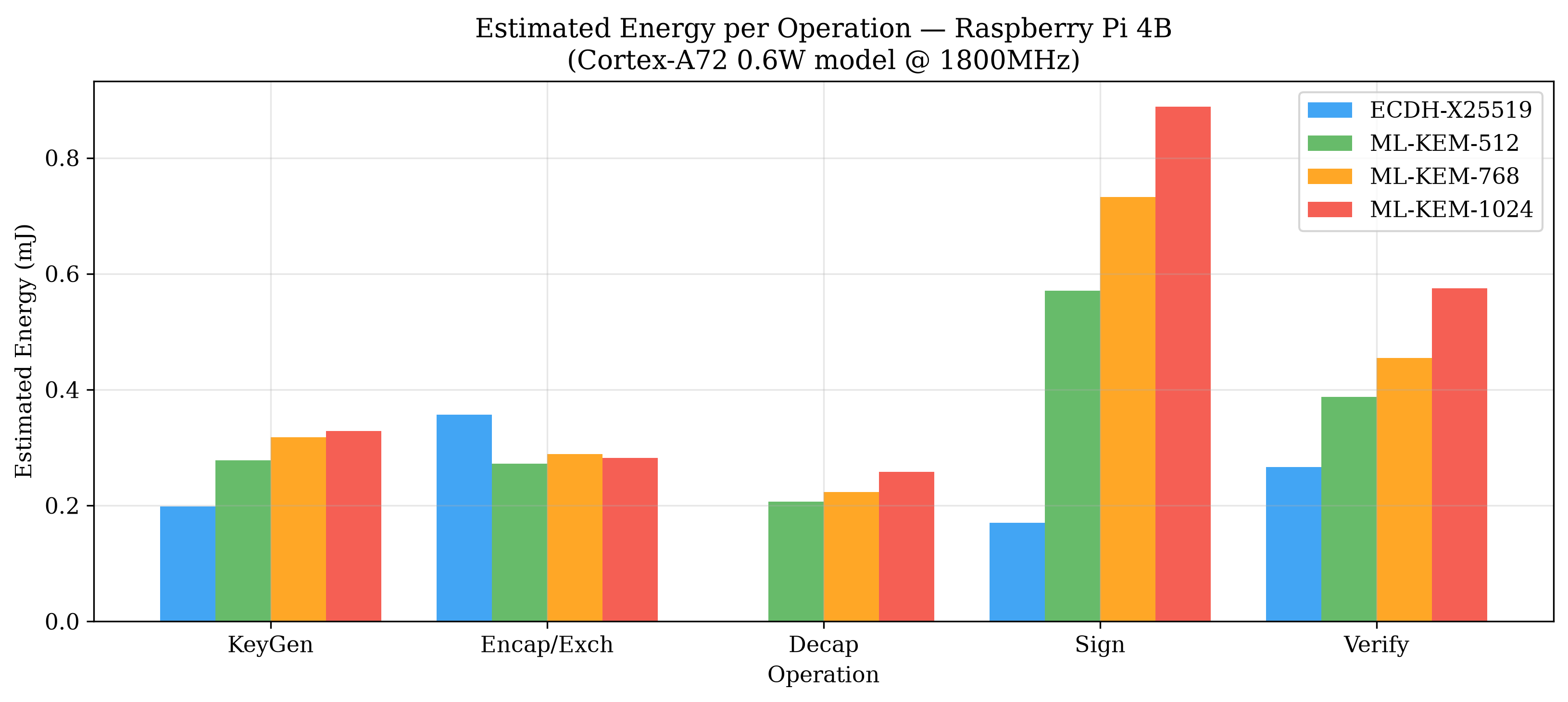}
\captionsetup{justification=centering}
\caption{Estimated energy per operation by configuration, corresponding to Table~\ref{tab:latency}.}
\label{fig:energy}
\end{figure*}

\subsection{Cryptographic Artifact Sizes}
\label{sec:results:sizes}

Table~\ref{tab:sizes} reports public key, ciphertext, shared secret, and signature sizes for each ML-KEM and ML-DSA security level. ML-KEM-768's 1{,}184-byte public key and 1{,}088-byte ciphertext, and ML-DSA-65's 3{,}309-byte signature, are the values transmitted in Steps 2, 3, and 5 of the handshake described in Section~\ref{sec:arch:flow}.

\begin{table*}[t]
\centering
\renewcommand{\arraystretch}{1.2}
\small
\begin{tabular*}{\textwidth}{@{\extracolsep{\fill}}lcccc@{}}
\toprule
\textbf{Config} & \textbf{PK (B)} & \textbf{CT (B)} & \textbf{SS (B)} & \textbf{Sig (B)} \\
\midrule
ML-KEM-512 $+$ ML-DSA-44 & 800 & 768 & 32 & 2420 \\
ML-KEM-768 $+$ ML-DSA-65 & 1184 & 1088 & 32 & 3309 \\
ML-KEM-1024 $+$ ML-DSA-87 & 1568 & 1568 & 32 & 4627 \\
\bottomrule
\end{tabular*}

\vspace{6pt}
\captionsetup{justification=centering}
\caption{Public key, ciphertext, shared secret, and signature sizes (bytes).}
\label{tab:sizes}
\end{table*}

\subsection{Full Handshake Latency}
\label{sec:results:handshake}

Table~\ref{tab:e2e} reports measured latency for the complete post-quantum handshake (Steps 1--6) and command round trip (Steps 7--9), against the classical ECDH baseline, over 500 runs. The ML-KEM-768 handshake completes in 2.48 ms on average, 0.38 ms slower than the 2.10 ms classical baseline.

\begin{table*}[t]
\centering
\renewcommand{\arraystretch}{1.2}
\small
\begin{tabular*}{\textwidth}{@{\extracolsep{\fill}}lccccc@{}}
\toprule
\textbf{Config} & \textbf{Phase} & \textbf{Mean (ms)} & \textbf{Std} & \textbf{P95} & \textbf{Energy (mJ)} \\
\midrule
ECDH baseline & Handshake (Steps~1--6) & 2.0987 & 0.4578 & 2.9963 & 1.259242 \\
 & Cmd encrypt & 0.0790 & 0.0286 & 0.1334 & 0.047379 \\
 & Cmd decrypt & 0.0369 & 0.0076 & 0.0496 & 0.022154 \\
\midrule
ML-KEM-768 proxy & Handshake (Steps~1--6) & 2.4765 & 0.7659 & 3.9971 & 1.485922 \\
 & Cmd encrypt & 0.0909 & 0.0398 & 0.1803 & 0.054524 \\
 & Cmd decrypt & 0.0427 & 0.0160 & 0.0703 & 0.025628 \\
\bottomrule
\end{tabular*}

\vspace{6pt}
\captionsetup{justification=centering}
\caption{Post-quantum handshake latency (Steps~1--6) and command round-trip (500 runs, long-term DSA keypair persists across sessions).}
\label{tab:e2e}
\end{table*}

\begin{figure}[t]
\centering
\includegraphics[width=0.85\linewidth]{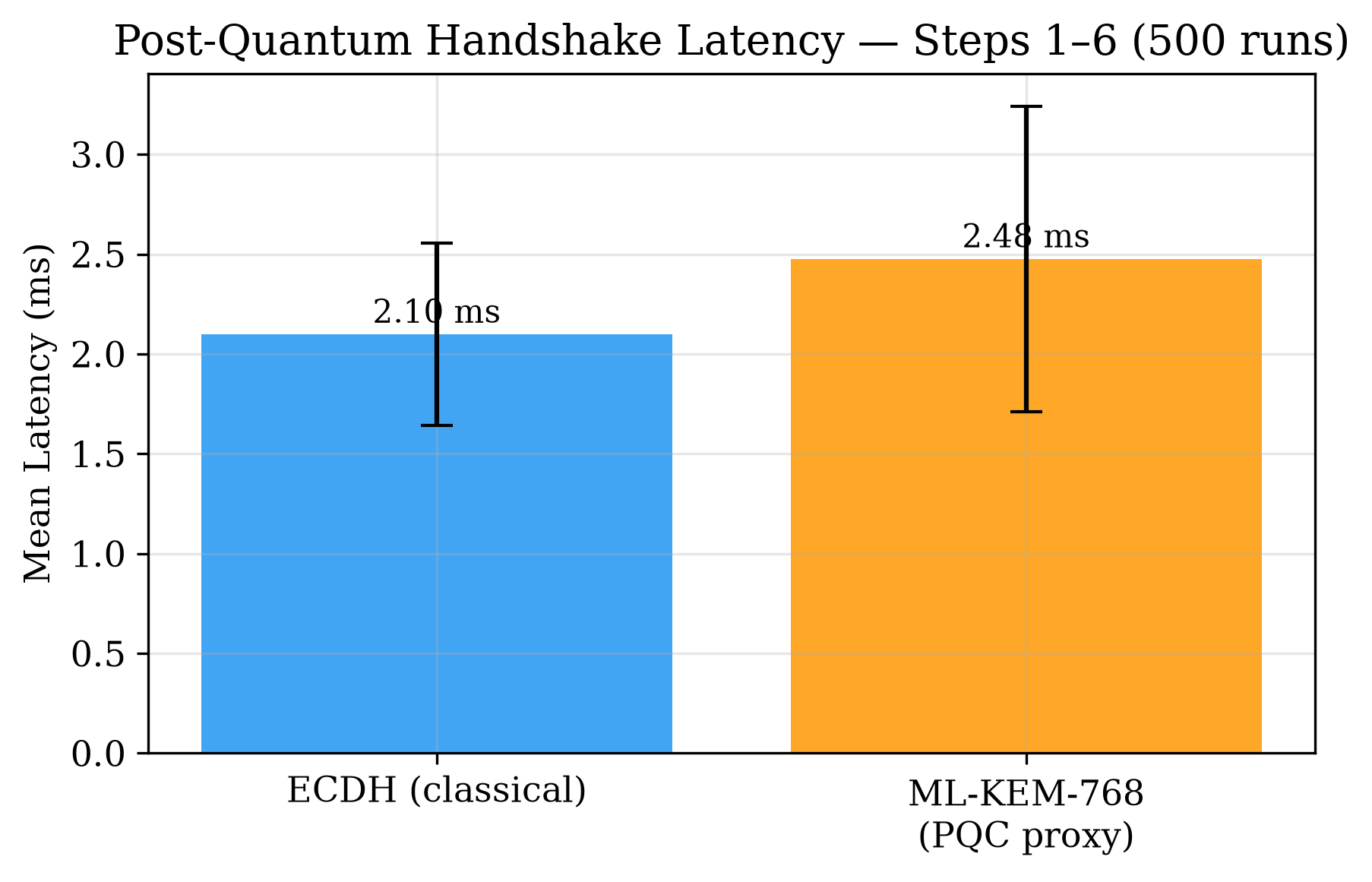}
\captionsetup{justification=centering}
\caption{Full handshake latency, ECDH baseline versus ML-KEM-768 proxy, corresponding to Table~\ref{tab:e2e}.}
\label{fig:handshakePlot}
\end{figure}

This result is worth comparing directly against the literature-derived estimate we reported at the conference stage of this work, before this empirical study was conducted. That earlier estimate, built from primitive-level timings reported by Fitzgibbon and Ottaviani~\cite{Fitzgibbon2024} rather than a full implementation of this architecture, projected a total handshake time of 8.04 ms. The measured value on real hardware, running the complete implementation described in Section~\ref{sec:implementation}, is 2.48 ms, roughly $3.2\times$ faster than the earlier estimate. This gap illustrates a general risk in projecting system-level latency from isolated primitive benchmarks: the estimate summed individual operation times without accounting for the actual scheduling, object construction, and measurement conditions of a running implementation, and in this case overstated the true cost substantially. It also means the architecture's performance case is stronger than what could be claimed prior to this empirical validation.

\subsection{Network Latency Impact}
\label{sec:results:network}

Table~\ref{tab:network} reports total handshake latency under simulated network round-trip times from 0 to 250 ms. At a 0 ms RTT, the measured total of 2.87 ms is close to the 2.48 ms handshake latency reported in Section~\ref{sec:results:handshake}, with the small difference attributable to the additional session setup overhead in the network benchmark's measurement loop. At any realistic RTT, network latency dominates total handshake time. At 20 ms RTT, representative of typical home broadband, PQC computation accounts for approximately 8\% of the 43.52 ms total.

\begin{table*}[t]
\centering
\renewcommand{\arraystretch}{1.2}
\small
\begin{tabular*}{\textwidth}{@{\extracolsep{\fill}}cccc@{}}
\toprule
\textbf{RTT (ms)} & \textbf{Mean (ms)} & \textbf{P95 (ms)} & \textbf{Energy (mJ)} \\
\midrule
0 & 2.870 & 4.340 & 1.721997 \\
20 & 43.517 & 44.655 & 26.109944 \\
50 & 103.599 & 104.588 & 62.159098 \\
100 & 203.628 & 204.989 & 122.177007 \\
250 & 503.705 & 504.704 & 302.223219 \\
\bottomrule
\end{tabular*}

\vspace{6pt}
\captionsetup{justification=centering}
\caption{ML-KEM-768 handshake latency under simulated network RTT (500 runs each, 4 one-way hops simulated).}
\label{tab:network}
\end{table*}

\begin{figure}[t]
\centering
\includegraphics[width=\linewidth]{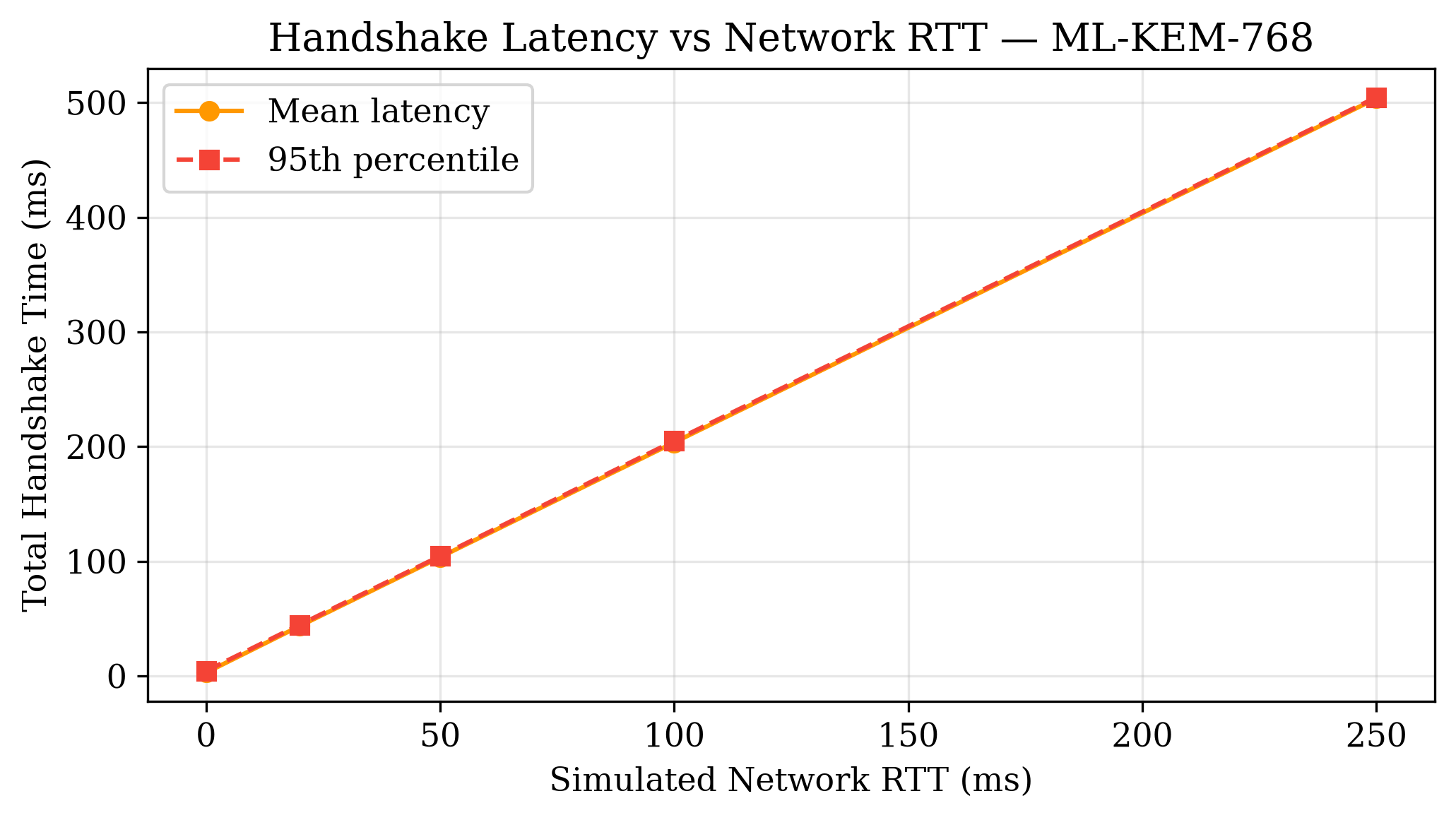}
\captionsetup{justification=centering}
\caption{Handshake latency versus simulated network round-trip time, corresponding to Table~\ref{tab:network}.}
\label{fig:network}
\end{figure}

\subsection{Energy Estimation}
\label{sec:results:energy}

Energy figures throughout this section follow the constant-power estimation model described in Section~\ref{sec:setup:energy}. The ML-KEM-768 handshake is estimated at 1.486 mJ, against 1.259 mJ for the classical baseline, a difference proportional to the measured latency gap reported in Section~\ref{sec:results:handshake}, since energy is computed as a fixed multiple of execution time under this model rather than measured independently. We report this consistently across every table in this section, but caution against treating the energy columns as an independent finding beyond the latency data they are derived from.

\subsection{Sequential Throughput and Scalability}
\label{sec:results:throughput}

Running 500 sequential sessions on the Raspberry Pi 4B, the proxy sustains 443.3 sessions per second, at a mean per-session latency of 2.22 ms, with 60 KB of memory growth and no leak detected, reported in full in Section~\ref{sec:results:memory}. Table~\ref{tab:extrapolation} extrapolates this measurement to larger deployment scenarios under the assumption of linear per-core scaling. A 32-core cloud instance is projected to sustain approximately 3{,}546 sessions per second under this assumption, though actual cloud throughput would additionally depend on network I/O and load balancing not modelled here.

\begin{table*}[t]
\centering
\renewcommand{\arraystretch}{1.2}
\small
\begin{tabular*}{0.65\textwidth}{@{\extracolsep{\fill}}lc@{}}
\toprule
\textbf{Deployment Scenario} & \textbf{Sessions / Second} \\
\midrule
Raspberry Pi~4B (1 core) & 110.8 \\
Raspberry Pi~4B (4 cores) & 443.3 \\
Cloud instance (8 cores) & 886.6 \\
Cloud instance (32 cores) & 3546.4 \\
\bottomrule
\end{tabular*}

\vspace{2pt}
\footnotesize
Linear scaling assumed. Actual cloud throughput depends on network I/O and load balancing.

\vspace{6pt}
\captionsetup{justification=centering}
\caption{Extrapolated ML-KEM-768 proxy throughput for deployment scenarios. Based on measured Pi~4B performance, assuming linear per-core scaling.}
\label{tab:extrapolation}
\end{table*}

\begin{figure}[t]
\centering
\includegraphics[width=\linewidth]{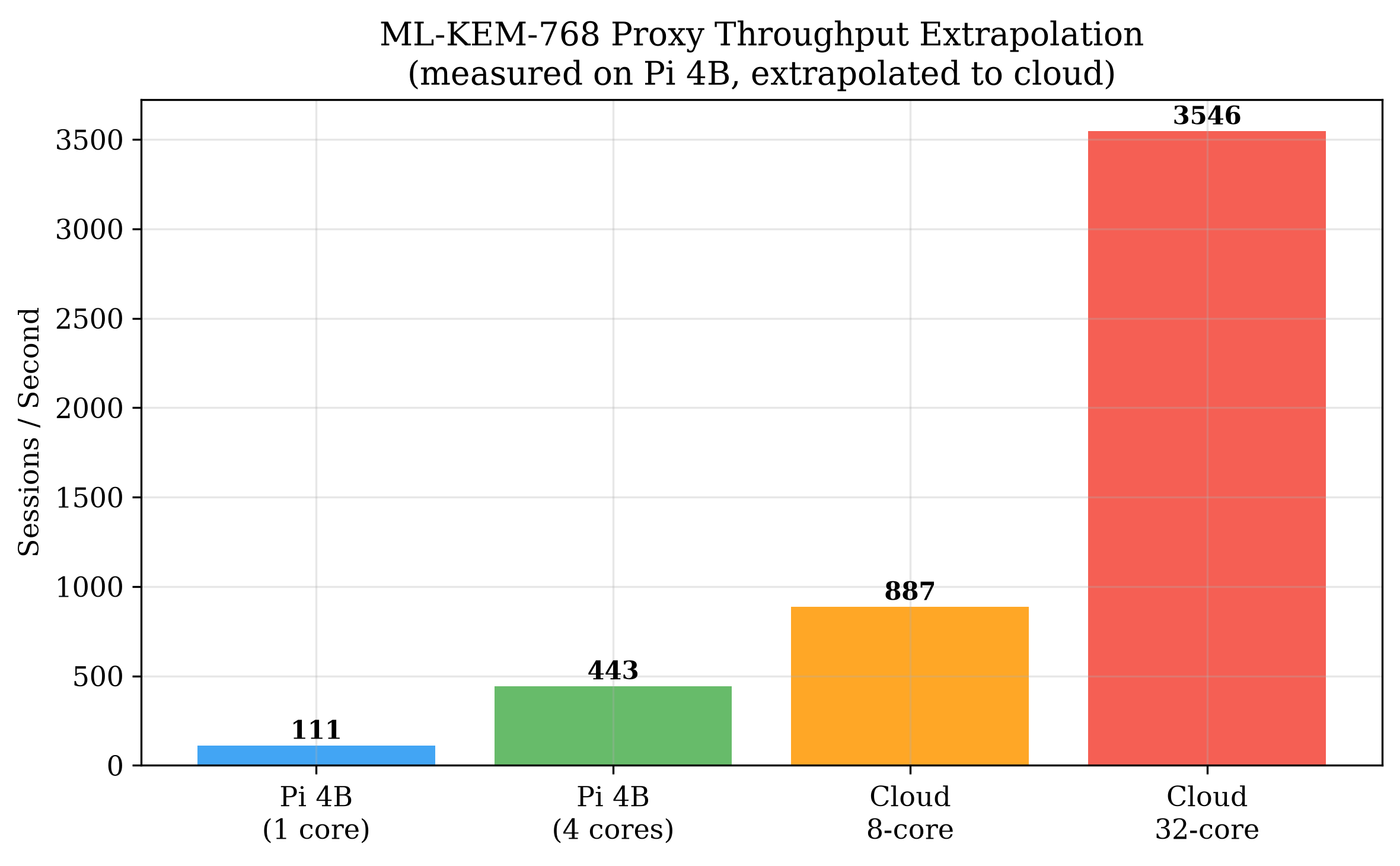}
\captionsetup{justification=centering}
\caption{Extrapolated throughput across deployment scenarios, corresponding to Table~\ref{tab:extrapolation}.}
\label{fig:extrapolation}
\end{figure}

\subsection{Concurrent Session Performance}
\label{sec:results:concurrent}

Table~\ref{tab:stress} reports throughput and latency under 1 to 32 simultaneous session threads. Every thread count achieves 100\% session success, with no failures at any concurrency level tested. Throughput peaks at 887.9 sessions per second at 4 concurrent threads, matching the Raspberry Pi 4B's 4 physical cores, and declines beyond that point as additional threads compete for the same 4 cores through context switching rather than genuine parallel execution. This is the expected pattern for a CPU-bound workload on fixed hardware, and it indicates that horizontal scaling across multiple gateway instances, rather than increasing thread count on a single 4-core device, is the appropriate strategy for serving more concurrent sessions than a single Pi 4B can handle in parallel.

\begin{table*}[t]
\centering
\renewcommand{\arraystretch}{1.2}
\small
\begin{tabular*}{\textwidth}{@{\extracolsep{\fill}}ccccc@{}}
\toprule
\textbf{Threads} & \textbf{Success (\%)} & \textbf{Mean lat. (ms)} & \textbf{Max lat. (ms)} & \textbf{Sessions/s} \\
\midrule
1 & 100.0 & 4.07 & 4.07 & 160.7 \\
2 & 100.0 & 3.34 & 3.43 & 473.6 \\
4 & 100.0 & 3.06 & 3.45 & 887.9 \\
8 & 100.0 & 7.41 & 10.32 & 358.2 \\
16 & 100.0 & 10.55 & 17.35 & 444.5 \\
32 & 100.0 & 13.92 & 23.87 & 303.6 \\
\bottomrule
\end{tabular*}

\vspace{2pt}
\footnotesize
Sequential throughput: 443.3 sessions/s over 500 sessions. Memory growth: 60.0~KB (no leak).

\vspace{6pt}
\captionsetup{justification=centering}
\caption{ML-KEM-768 proxy performance under concurrent load (Raspberry Pi~4B).}
\label{tab:stress}
\end{table*}

\begin{figure*}[t]
\centering
\includegraphics[width=1.0\textwidth]{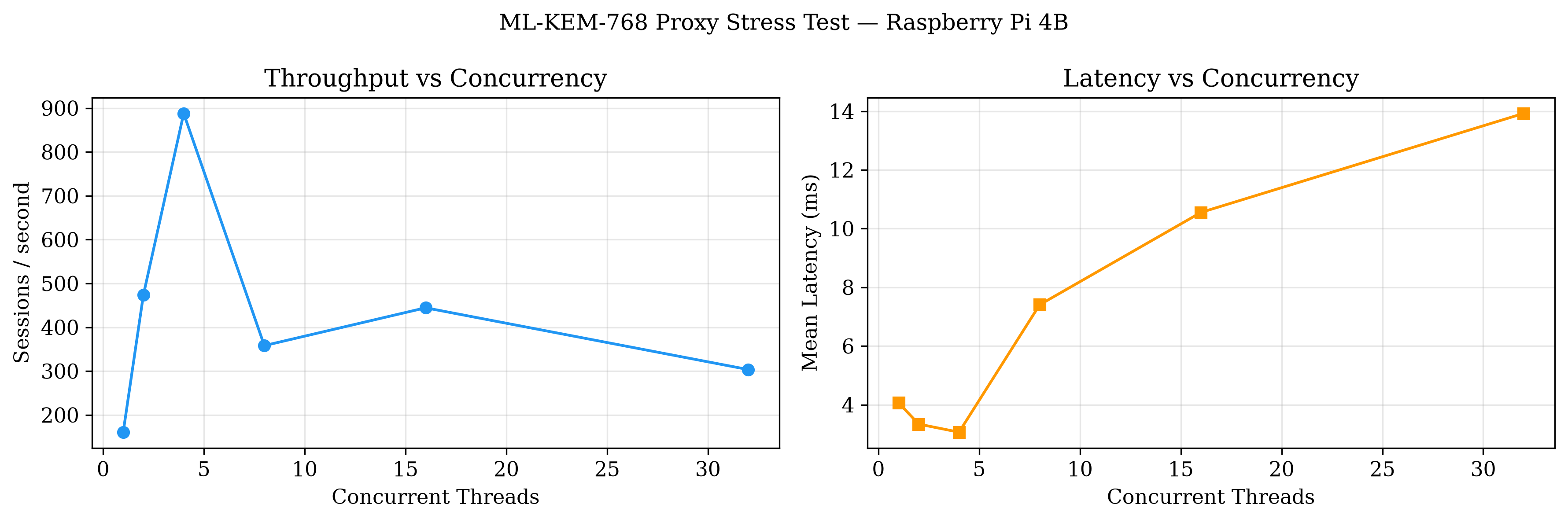}
\captionsetup{justification=centering}
\caption{Throughput and latency versus concurrent thread count, corresponding to Table~\ref{tab:stress}.}
\label{fig:concurrent}
\end{figure*}

\subsection{Memory Stability Under Load}
\label{sec:results:memory}

Process memory grew by 60 KB over 500 sequential sessions, from 31{,}942 KB to 32{,}002 KB, well below the 5 MB growth threshold used to flag a potential leak in this study. No memory leak is detected under sustained sequential load. This result depends on the session key zeroization described in Section~\ref{sec:impl:proxy}, since a session key that accumulated in memory across hundreds of sessions without being cleared would be expected to produce measurable growth at this sample size.

\begin{figure}[t]
\centering
\includegraphics[width=0.85\linewidth]{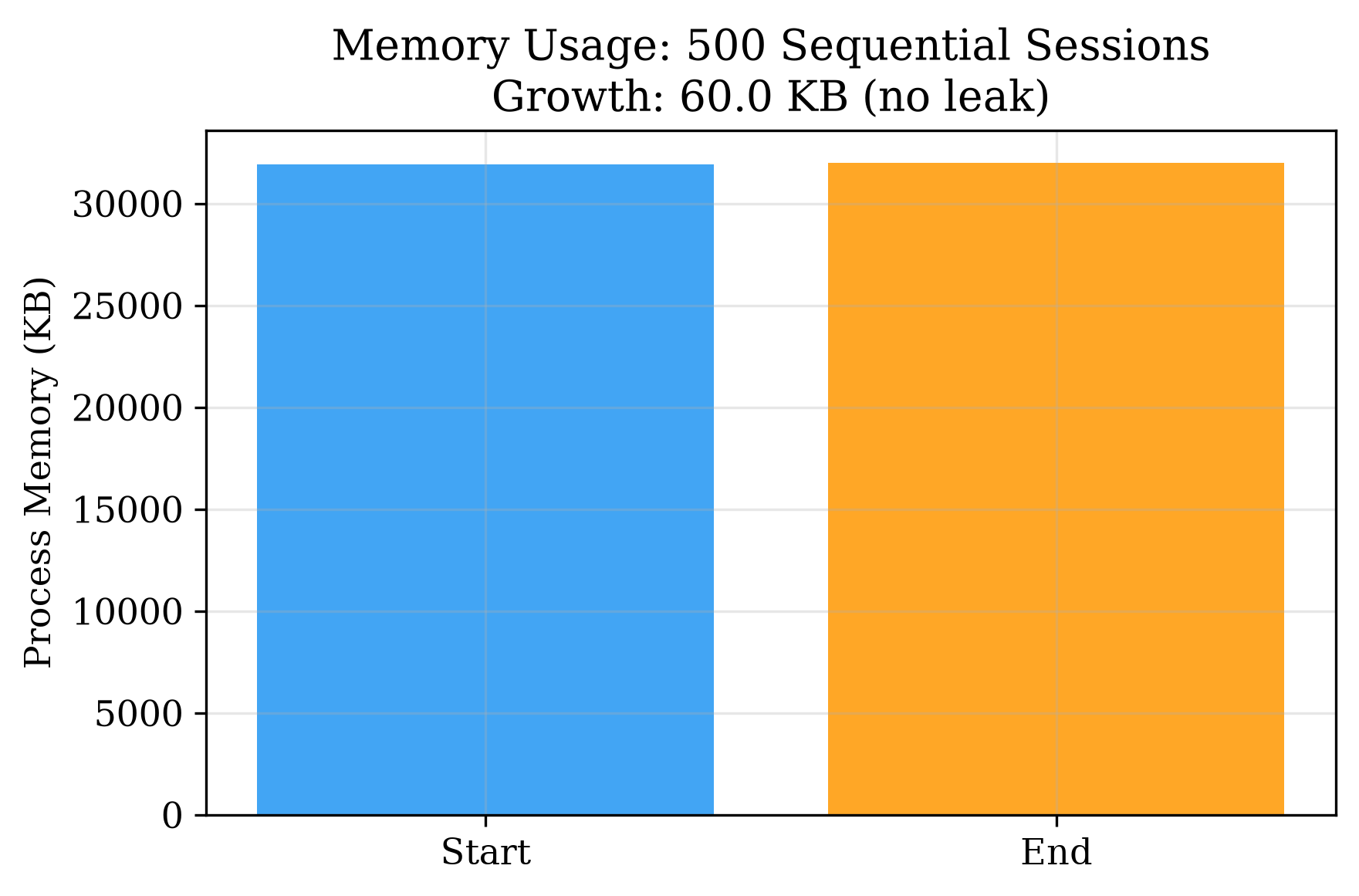}
\captionsetup{justification=centering}
\caption{Process memory at the start and end of 500 sequential sessions.}
\label{fig:memstability}
\end{figure}

\subsection{ESP32-S3 Device-Side Infeasibility}
\label{sec:results:esp32}

Table~\ref{tab:esp32} summarises the ESP32-S3 measurements introduced in Section~\ref{sec:background:hvac} and described methodologically in Section~\ref{sec:impl:esp32}. The device has 339 KB of free heap at runtime, out of 380 KB total SRAM. ML-KEM-768 key generation requires 900 KB of working memory, and a direct \texttt{malloc(900\,KB)} call fails. The same device successfully runs AES-128 encryption in 0.032 ms and ECDH-P256 key generation in 111.93 ms, confirming that the infeasibility is specific to ML-KEM-768's memory requirement rather than a general inability to perform cryptographic operations, and that even the classical asymmetric operation it can run is already too slow to be practical on this hardware class.

\begin{table*}[t]
\centering
\renewcommand{\arraystretch}{1.2}
\small
\begin{tabular*}{\textwidth}{@{\extracolsep{\fill}}ll@{}ll@{}ll@{}}
\toprule
\multicolumn{2}{c}{\textbf{Hardware}} &
\multicolumn{2}{c}{\textbf{Classical Cryptography}} &
\multicolumn{2}{c}{\textbf{Post-Quantum Cryptography}} \\
\cmidrule(lr){1-2}
\cmidrule(lr){3-4}
\cmidrule(lr){5-6}
\textbf{Metric} & \textbf{Value} &
\textbf{Operation} & \textbf{Latency} &
\textbf{Metric} & \textbf{Value} \\
\midrule
Platform & ESP32-S3 &
AES-128 & 0.0320~ms &
ML-KEM-768 memory & 900~KB \\
CPU frequency & 240~MHz &
ECDH-P256 & 111.9270~ms &
malloc(900~KB) & \textbf{FAILED} \\
Total SRAM & 380~KB &
&
&
ML-KEM-768 feasible & \textbf{No} \\
Free heap & 339~KB &
&
&
&
\\
\bottomrule
\end{tabular*}

\vspace{6pt}
\captionsetup{justification=centering}
\caption{ESP32-S3 memory feasibility for ML-KEM-768 and classical cryptography timing, including classical timing results and memory allocation failure.}
\label{tab:esp32}
\end{table*}

\begin{figure}[t]
\centering
\includegraphics[width=\linewidth]{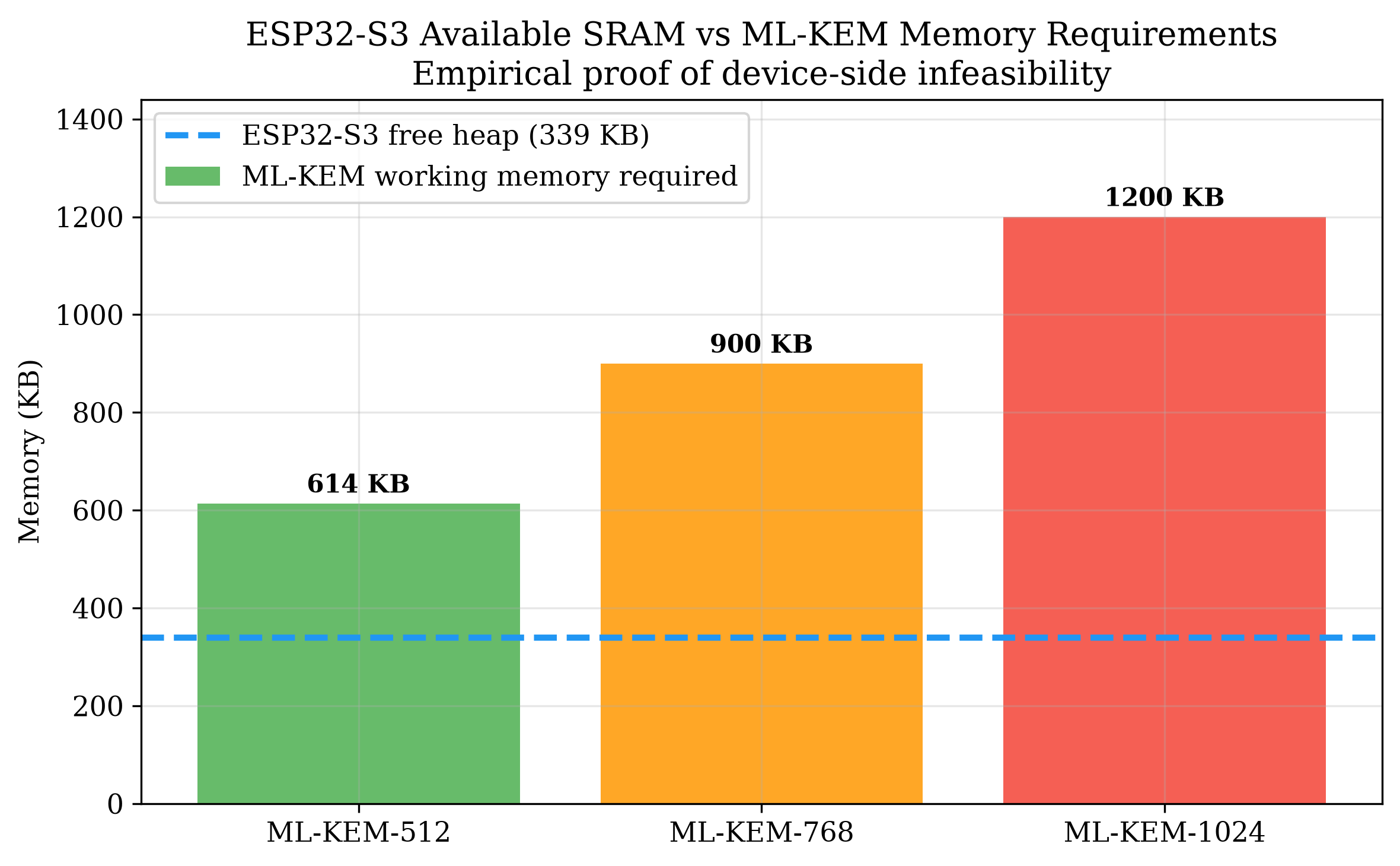}
\captionsetup{justification=centering}
\caption{ESP32-S3 free heap versus ML-KEM working memory requirements by security level, corresponding to Table~\ref{tab:esp32}.}
\label{fig:esp32mem}
\end{figure}

\section{Security Analysis}
\label{sec:security}

\subsection{Threat Model}
\label{sec:security:threat}

We consider a quantum-capable adversary with access to a cryptographically relevant quantum computer, capable of executing Shor's algorithm~\cite{Shor1997} and Grover's algorithm~\cite{Grover1996} at scale. The adversary can passively observe and record all network traffic on both the mobile application to PQC Layer link and the PQC Layer to vendor cloud link, consistent with the Harvest Now, Decrypt Later collection model described in Section~\ref{sec:background:threat}. The adversary can additionally perform active attacks on the mobile application to PQC Layer link, including man in the middle interception and message substitution. The adversary does not have physical access to the PQC Layer server, the mobile application device, or the HVAC controller, and cannot break the Module-LWE problem on which both ML-KEM and ML-DSA security reduce~\cite{FIPS203,FIPS204,Peikert2019}.

Side channel attacks against this implementation, evaluated empirically in Section~\ref{sec:security:sidechannel}, physical tampering, and compromise of the PQC Layer's private keys fall partly within and partly outside this network level threat model, as detailed in Section~\ref{sec:security:limitations}.

\subsection{Security Properties}
\label{sec:security:properties}

Table~\ref{tab:secprops} summarises the security properties provided by the architecture described in Section~\ref{sec:architecture}, each following directly from the protocol steps in Section~\ref{sec:arch:flow}.

\begin{table*}[t]
\centering
\renewcommand{\arraystretch}{1.2}
\small
\setlength{\arrayrulewidth}{0.1pt}

\begin{tabularx}{\textwidth}{
    @{}
    >{\hsize=0.9\hsize\raggedright\arraybackslash}X
    >{\hsize=0.9\hsize\raggedright\arraybackslash}X
    >{\hsize=1.2\hsize\raggedright\arraybackslash}X
    @{}
}
\toprule
\textbf{Property} & \textbf{Mechanism} & \textbf{Guarantee} \\
\midrule

Quantum-safe key establishment
& ML-KEM-768~\cite{FIPS203}
& Shared secret protected under Module-LWE \\
\cline{1-3}

Per-session forward secrecy
& Fresh keypair per session~\cite{FIPS203}
& Past sessions remain unrecoverable \\
\cline{1-3}

Entity authentication
& ML-DSA-65~\cite{FIPS204}
& PQC Layer identity authenticated \\
\cline{1-3}

Message confidentiality
& AES-256-GCM~\cite{McGrew2004}
& Commands remain confidential \\
\cline{1-3}

Message integrity
& AES-256-GCM GHASH tag~\cite{McGrew2004}
& Ciphertext modification detected \\
\cline{1-3}

Replay resistance
& GCM tag $+$ per-session nonce registry
& Replay attempts detected and rejected \\
\cline{1-3}

Non-invasive boundary
& No device modification
& HVAC attack surface remains unchanged \\

\bottomrule
\end{tabularx}

\vspace{6pt}
\captionsetup{justification=centering}
\caption{Security properties of the proposed architecture.}
\label{tab:secprops}
\end{table*}

Entity authentication merits particular attention, since it is what prevents the man in the middle attack evaluated empirically in Section~\ref{sec:security:sidechannel}. Without ML-DSA-65 authentication, an adversary could intercept the ML-KEM public key transmitted in Step 2, substitute their own key, and establish separate sessions with both the mobile application and the legitimate PQC Layer. ML-DSA-65 prevents this by binding the handshake transcript, comprising the Step 1 nonce, the ML-KEM public key, and the ciphertext, to the PQC Layer's long-term signing identity, so that an adversary without the PQC Layer's private signing key cannot produce a transcript the mobile application will accept.

\subsection{Practical Side-Channel Analysis}
\label{sec:security:sidechannel}

We evaluate four empirical side-channel properties of the implementation described in Section~\ref{sec:implementation}, plus one descriptive timing comparison, all measured on the architecture's actual decapsulation path, meaning \texttt{PQCProxy.decapsulate()} rather than the underlying \texttt{liboqs} primitive in isolation. Table~\ref{tab:sidechannel} summarises the four pass or fail tests. All four pass.

\begin{table*}[t]
\centering
\renewcommand{\arraystretch}{1.2}
\small

\begin{tabularx}{\textwidth}{
    @{}
    >{\hsize=1\hsize\raggedright\arraybackslash}X
    >{\hsize=1.8\hsize\raggedright\arraybackslash}X
    >{\hsize=0.2\hsize\centering\arraybackslash}X
    @{}
}
\toprule
\textbf{Test} & \textbf{Key Result} & \textbf{Status} \\
\midrule
Memory Forensics
& Session key not found post-zeroization
& \checkmark \\

MITM Attack Simulation
& ML-DSA-65 raised ValueError, forgery blocked
& \checkmark \\

Nonce Collision Test
& 10{,}000/10{,}000 unique, $P = 6.310\times10^{-22}$
& \checkmark \\

TVLA Fixed-vs-Random
& $t = -0.268$ (threshold $\pm4.5$)
& \checkmark \\

\bottomrule
\end{tabularx}

\vspace{6pt}
\captionsetup{justification=centering}
\caption{Side-channel analysis results. Memory forensics, MITM, nonce, and TVLA tests follow documented methodologies; AES-GCM timing comparison in Table~\ref{tab:cvbaseline} is descriptive.}
\label{tab:sidechannel}
\end{table*}

Memory forensics verifies the in-place session key zeroization described in Section~\ref{sec:impl:proxy}, by inspecting the exact memory address of the session key buffer before and after \texttt{end\_session()} rather than searching process memory broadly for a copy of the key, which would remain trivially findable as long as the search itself holds a live reference to the value it is searching for. No nonzero bytes remain at that address after session termination.

The MITM simulation isolates two claims separately: whether an attacker who substitutes their own ML-KEM public key can recover the resulting shared secret, and whether that attacker can subsequently forge a valid ML-DSA-65 signature over the substituted transcript to get the victim to accept the session. The attacker succeeds at the first and fails at the second, raising \texttt{ValueError} during signature verification, which is the intended behaviour described in Section~\ref{sec:security:properties}.

The nonce collision test generates 10{,}000 random 96-bit nonces with zero collisions, consistent with the theoretical collision probability of $6.310\times10^{-22}$ under the birthday bound, and confirms the replay resistance mechanism described in Section~\ref{sec:arch:flow} is not undermined by nonce reuse in practice.

The TVLA result in Table~\ref{tab:sidechannel} uses the standard fixed-versus-random Welch's t-test methodology~\cite{Goodwill2011}, comparing decapsulation timing on a repeatedly reused ciphertext against a freshly generated ciphertext each iteration, with all random-group inputs pre-generated before timing begins to avoid confounding the comparison with object construction overhead. At $t = -0.268$, well inside the $\pm4.5$ threshold, no timing leakage is detected between fixed and random inputs on the architecture's decapsulation path.

Table~\ref{tab:cvbaseline} reports timing variability, measured as coefficient of variation, on the same decapsulation path relative to AES-256-GCM, an operation already used elsewhere in this architecture and a widely accepted constant-time primitive, measured in the same run on the same hardware. We report this as a descriptive noise-floor comparison rather than a pass or fail test, since no universal coefficient of variation threshold is established in the literature independent of the measurement environment.

\begin{table*}[t]
\centering
\renewcommand{\arraystretch}{1.2}
\small
\begin{tabular*}{\textwidth}{@{\extracolsep{\fill}}lcc@{}}
\toprule
\textbf{Operation} & \textbf{Mean (ms)} & \textbf{CV (\%)} \\
\midrule
AES-256-GCM (baseline) & 0.1377 & 80.68 \\
PQCProxy.decapsulate() & 1.2326 & 38.80 \\
\bottomrule
\end{tabular*}

\vspace{2pt}
\footnotesize
Difference: $-41.88$ percentage points relative to baseline.
\vspace{6pt}
\captionsetup{justification=centering}
\caption{Timing variability of the decapsulation path and AES-256-GCM baseline (5{,}000 samples each). Descriptive comparison, not a pass/fail test.}
\label{tab:cvbaseline}
\end{table*}

\begin{figure*}[t]
\centering
\includegraphics[width=\textwidth]{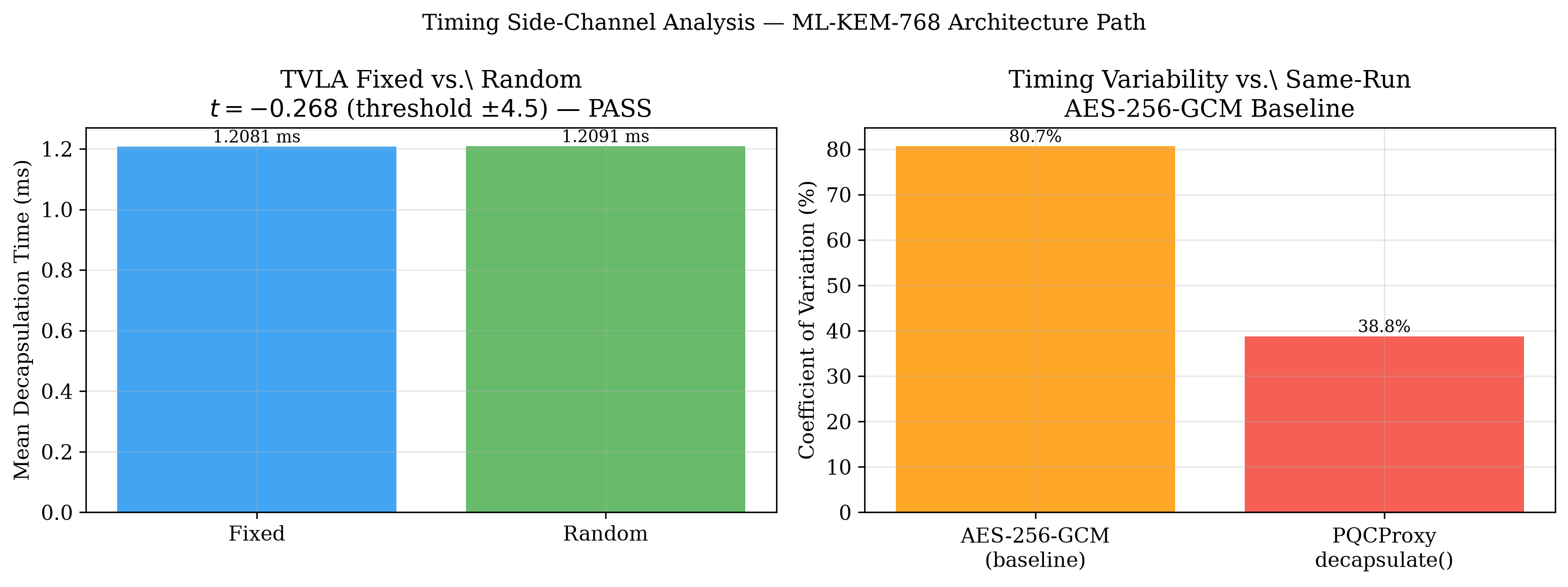}
\captionsetup{justification=centering}
\caption{TVLA fixed-versus-random decapsulation timing and coefficient of variation relative to the AES-256-GCM baseline, corresponding to Table~\ref{tab:sidechannel} and Table~\ref{tab:cvbaseline}.}
\label{fig:sidechanneltiming}
\end{figure*}

The architecture's decapsulation path shows a lower coefficient of variation than the AES-256-GCM baseline measured in the same run, which is a useful sanity check but not by itself a security claim. Both operations run on a general-purpose Linux kernel sharing the CPU with the scheduler, network stack, and other system processes, so absolute coefficient of variation values at this timescale reflect operating system level timing noise as much as algorithmic behaviour. The TVLA result above, not this table, is the methodology we rely on for an actual leakage determination, consistent with its design purpose.

This analysis addresses timing side channels only. Power and electromagnetic side-channel analysis require dedicated instrumentation, an oscilloscope and near-field probes capturing hardware level power or emission traces, that was not available for this study, and we identify it explicitly as future work in Section~\ref{sec:limitations} rather than a gap left unacknowledged.

\subsection{Acknowledged Limitations}
\label{sec:security:limitations}

The most significant architectural limitation is the trusted proxy assumption. The PQC Layer terminates the quantum-safe session and holds decrypted plaintext commands before re-encrypting them for the vendor cloud, described in Step 8a of Section~\ref{sec:arch:flow}. An adversary who compromises the PQC Layer itself, as opposed to the network links it mediates, gains access to plaintext HVAC commands. This reduces the system's security to the security of the PQC Layer rather than providing end-to-end cryptographic protection all the way to the HVAC device, a trade-off inherent to the non-invasive design requirement established in Section~\ref{sec:arch:principles} and consistent with the trusted proxy model used in related IoT gateway literature~\cite{Basile2024,Lin2016}. This trust model is not new relative to the existing deployment, since the vendor cloud already holds the same plaintext commands and behavioural data today, with no user oversight, under the classical architecture described in Section~\ref{sec:background:hvac}.

The link between the PQC Layer and the vendor cloud remains classical TLS, as established in Section~\ref{sec:arch:overview}. Full end-to-end quantum safety would require the vendor cloud to support ML-KEM natively, which is outside the scope of a non-invasive migration strategy and requires vendor cooperation this architecture is explicitly designed not to depend on.

The architecture does not provide mutual authentication between the PQC Layer and the HVAC device itself. The device authenticates to the vendor cloud using its existing credentials, and no post-quantum device identity mechanism is introduced. Device-side authentication and credential rotation for legacy HVAC devices are deferred to future work.

The PQC Layer introduces a new network node and a potential single point of failure, acknowledged in Section~\ref{sec:arch:justification}. Section~\ref{sec:results:concurrent} reports the throughput and latency behaviour of a single instance under concurrent load, but production deployment would additionally require high-availability configuration, load balancing, and geographic redundancy not evaluated in this study.

\section{Discussion}
\label{sec:discussion}

\subsection{Performance vs Prior Estimates}
\label{sec:disc:performance}

The gap between our conference-stage estimate and the measured result reported in Section~\ref{sec:results:handshake} is worth discussing on its own terms, beyond the headline comparison. The earlier estimate of 8.04 ms was built by summing individual primitive-level timings reported by Fitzgibbon and Ottaviani~\cite{Fitzgibbon2024} for ML-KEM-768 keygen, encapsulation, and decapsulation, and ML-DSA-65 signing and verification, measured in isolation on comparable hardware. The measured value on our own implementation, running the complete protocol described in Section~\ref{sec:arch:flow} rather than isolated primitive calls, is 2.48 ms, roughly a third of the earlier estimate.

This is not a case of the earlier literature being wrong. Fitzgibbon and Ottaviani's numbers, when we independently sum the corresponding primitive latencies from our own Table~\ref{tab:latency}, keygen, encap, decap, and sign for ML-KEM-768 and ML-DSA-65, come to approximately 2.98 ms, which is close to our own measured full handshake time of 2.48 ms. The 8.04 ms conference-stage estimate overstated the true cost, most likely because it did not account for the specific measurement conditions, warm caches, persistent long-term keys, and connection reuse, that a running proxy implementation actually benefits from across a full session. The lesson generalises beyond this paper: projecting system-level latency by summing isolated primitive benchmarks is a reasonable first approximation for a proposal-stage architecture paper, but it is not a substitute for measuring the full implementation, and the direction of the error is not guaranteed to be conservative. In our case the estimate was pessimistic, which happened to strengthen the eventual case for the architecture, but a researcher relying on this style of estimate to argue feasibility should treat it as provisional until an implementation confirms it.

\subsection{The ECDH-on-Device Finding}
\label{sec:disc:ecdh}

Section~\ref{sec:results:esp32} reports that ECDH-P256 key generation on the ESP32-S3 takes 111.93 ms, roughly two orders of magnitude slower than the same operation on the Raspberry Pi 4B reported in Section~\ref{sec:results:operations}. This result is worth separating from the ML-KEM-768 memory infeasibility finding it sits alongside, because it points to a different and in some ways more fundamental problem. The memory infeasibility result is specific to post-quantum algorithms, ML-KEM-768 needs 900 KB and the device has 339 KB free. The ECDH result is not. It shows that even the classical asymmetric cryptography this device class is currently expected to run is already too slow to be comfortable in an interactive HVAC control context, independent of any quantum consideration at all.

This reframes the motivation for a proxy architecture slightly. The case for moving asymmetric cryptography off the HVAC device is not only a forward-looking response to the quantum threat described in Section~\ref{sec:background:threat}, it is also a response to a present-day performance limitation on the device class this architecture targets. A migration strategy that only addresses the post-quantum transition, while leaving classical asymmetric operations running slowly on constrained device firmware, would still leave a real problem unsolved. The architecture proposed in this paper offloads all asymmetric cryptography, not only the post-quantum portion of it, which this finding suggests is the more complete fix for this hardware class.

\subsection{Deployment Considerations}
\label{sec:disc:deployment}

Section~\ref{sec:arch:overview} describes three physical deployment models for the PQC Layer: a cloud-hosted proxy operated by a third party, a home gateway running on user-owned hardware such as the Raspberry Pi 4B used throughout this study, and eventual vendor-native integration that would make the proxy unnecessary. The throughput and scalability results in Section~\ref{sec:results:throughput} and Section~\ref{sec:results:concurrent} are directly relevant to choosing between the first two.

A single Raspberry Pi 4B sustains 443.3 sessions per second and handles 32 concurrent sessions with 100\% success, which is far beyond what a single household needs, a home gateway deployment has substantial headroom even on inexpensive hardware. For a cloud-hosted deployment serving many households from shared infrastructure, the extrapolation in Table~\ref{tab:extrapolation} suggests horizontal scaling across multiple gateway-class instances, consistent with the concurrency behaviour discussed in Section~\ref{sec:results:concurrent}, where throughput peaked at the Raspberry Pi 4B's own core count and did not improve further from additional threads on the same device. This supports scaling out across instances rather than up within a single instance, standard practice for cloud proxy services and consistent with the high-availability recommendation in Section~\ref{sec:security:limitations}.

\subsection{Migration Strategy and Regulatory Alignment}
\label{sec:disc:migration}

The deprecation and disallowance deadlines set out in NIST IR 8547~\cite{NISTIR8547}, described in Section~\ref{sec:background:threat}, apply directly to federal systems, but they signal the compliance direction commercial vendors are likely to follow on a similar timeline, alongside comparable product-security and lifecycle-update requirements under the EU Cyber Resilience Act~\cite{CRA2024}. Neither instrument requires or currently addresses legacy consumer HVAC devices specifically, which is precisely the gap this architecture targets. Homeowners and installers cannot wait for HVAC manufacturers to redesign device firmware or renegotiate cloud infrastructure on a regulatory timeline they do not control, and the device-side infeasibility results in Section~\ref{sec:results:esp32} confirm that firmware-level migration is not a near-term option for hardware already deployed.

The architecture's positioning as an interim, non-invasive layer, established in Section~\ref{sec:arch:justification} and reinforced by the cryptographic agility principle in Section~\ref{sec:arch:principles}, is a direct response to this mismatch between regulatory timelines and device replacement cycles. It provides a migration path that does not depend on vendor cooperation or coordinated regulatory enforcement reaching consumer IoT, while remaining structurally ready to be decommissioned, per vendor ecosystem, once that vendor eventually does adopt ML-KEM natively, without requiring any change to the HVAC device or the user's mobile application at that point.

\section{Limitations and Future Work}
\label{sec:limitations}

\subsection{Methodological Limitations}
\label{sec:limitations:methodological}

Several limitations of this study are methodological rather than architectural, and are distinct from the trusted proxy and single point of failure limitations already discussed in Section~\ref{sec:security:limitations}.

Energy figures throughout this paper follow the constant-power estimation model described in Section~\ref{sec:setup:energy}, not direct hardware measurement. This model captures relative energy cost between operations of different duration, but cannot detect any genuine difference in power draw between operations of similar duration, since energy is computed as a fixed multiple of execution time in every case. Direct hardware instrumentation, an in-line current sensor on the Raspberry Pi 4B's power supply, would allow energy to be measured rather than estimated, and is the most direct extension of the energy results in Section~\ref{sec:results:energy}.

The side-channel analysis in Section~\ref{sec:security:sidechannel} addresses timing side channels only. Power and electromagnetic side-channel analysis require dedicated instrumentation, an oscilloscope and near-field probes capturing hardware-level power or emission traces during cryptographic operations, that was not available for this study. A definitive constant-time determination for this implementation would require that instrumentation, applied to the same decapsulation path evaluated here.

The ESP32-S3 measurements reported in Section~\ref{sec:results:esp32} establish infeasibility through a direct memory allocation test and separately timed classical operations, rather than a full \texttt{liboqs} execution attempt on the device itself, as described in Section~\ref{sec:impl:esp32}. This is sufficient to demonstrate that ML-KEM-768 key generation cannot allocate the memory it requires, but it does not characterise how far into key generation the device would proceed before failing, or whether a partial, memory-optimised implementation might narrow the gap.

All benchmarks in this study run on a single Raspberry Pi 4B and a single ESP32-S3 development board. Results on other ARM-class gateway hardware, or on x86-based home server hardware increasingly used for self-hosted infrastructure, are not evaluated. The stress testing in Section~\ref{sec:results:concurrent} covers 500 sequential sessions and up to 32 concurrent threads, on the order of minutes of sustained load, and does not evaluate multi-day or multi-week stability, where more gradual memory or performance drift might appear that the sample sizes used here would not detect.

Finally, this study evaluates the mobile application to PQC Layer link and the PQC Layer's local behaviour directly, but the PQC Layer to vendor cloud link, Steps 8b and 8c in Section~\ref{sec:arch:flow}, is represented by the network latency simulation in Section~\ref{sec:results:network} rather than a live connection to an actual commercial vendor cloud. This follows directly from the non-invasive design constraint established in Section~\ref{sec:arch:principles}, since commercial HVAC vendor infrastructure is proprietary and inaccessible for independent academic testing, but it means the vendor-facing segment of the architecture is validated by simulation rather than live integration.

\subsection{Future Work}
\label{sec:limitations:future}

Beyond the direct hardware energy and power side-channel instrumentation already identified above, several extensions follow naturally from this work. Hybrid key exchange, combining ML-KEM-768 with a classical algorithm such as X25519 in a single handshake, would provide defence in depth during the migration period, at the cost of the additional key material and computation Section~\ref{sec:results:operations} would need to re-measure for that combined configuration. Device-side authentication and credential rotation for the HVAC controller itself, acknowledged as unaddressed in Section~\ref{sec:security:limitations}, would close the one link in the architecture that still relies entirely on the vendor cloud's existing device identity mechanism. Formal verification of the authentication and key establishment flow in Section~\ref{sec:arch:flow}, using a protocol verification tool such as ProVerif or Tamarin, would complement the empirical security analysis in Section~\ref{sec:security} with a machine-checked proof of the properties summarised in Table~\ref{tab:secprops}. Finally, a production-oriented reimplementation in a compiled language, rather than the Python implementation described in Section~\ref{sec:impl:stack}, would characterise how much of the measured latency in Section~\ref{sec:results:handshake} is attributable to the cryptographic operations themselves versus interpreter and object-construction overhead, and would represent the performance profile of an eventual deployed system more directly.

\section{Conclusions}
\label{sec:conclusion}

This paper presented a non-invasive, cloud-based architecture for integrating post-quantum cryptography into smart home HVAC systems, and evaluated it empirically on real hardware rather than through estimation alone. Legacy HVAC controllers cannot run ML-KEM directly. We confirmed this on an actual ESP32-S3 device, where a 900 KB memory requirement fails against 339 KB of available heap, and where even the classical ECDH operation this hardware class currently performs already takes 111.93 ms, too slow to be comfortable in an interactive control context independent of any quantum consideration. The PQC proxy layer proposed to address this, inserted between the mobile application and the vendor cloud, requires no change to the HVAC device, its firmware, or the vendor's existing infrastructure.

Measured on a Raspberry Pi 4B over 500 runs, the complete post-quantum handshake, ML-KEM-768 key encapsulation and ML-DSA-65 authentication, adds 0.38 ms over a classical baseline, a result roughly three times faster than the literature-derived estimate reported at the conference stage of this work, and a reminder that summing isolated primitive benchmarks is not a substitute for measuring a full implementation. The architecture sustains 443 sessions per second on commodity hardware with no memory leak detected under sustained load, and four independent side-channel tests, including a fixed-versus-random TVLA timing analysis on the architecture's actual decapsulation path, found no exploitable timing leakage, verified in-place session key zeroization, and confirmed that ML-DSA-65 blocks man-in-the-middle impersonation.

The trusted proxy assumption remains the architecture's most significant limitation, and is not new relative to the vendor cloud dependency the existing classical deployment already carries. The architecture is explicitly framed as an interim migration strategy rather than a permanent end state, deployable today without vendor cooperation, and structurally ready to be decommissioned once HVAC vendors adopt NIST's post-quantum standards natively.

\vspace{6pt}
\authorcontributions{Conceptualization, Mahedee M. and Kaysarul A.; methodology, Mahedee M.; software, Mahedee M. and Kaysarul A.; validation, Mahedee M., Kaysarul A. and Hasibul H.; formal analysis, Mahedee M.; investigation, Mahedee M.; writing, original draft preparation, Mahedee M.; writing, review and editing, Kaysarul A., Hasibul H. and SK Mizanur R.; supervision, SK Mizanur R. All authors have read and agreed to the published version of the manuscript.}

\funding{This research received no external funding.}

\institutionalreview{Not applicable.}

\informedconsent{Not applicable.}

\dataavailability{The benchmark source code, raw JSON results, and reproduction instructions are available from the corresponding author upon request and will be published in a public repository concurrent with journal submission.}

\acknowledgments{The authors thank the Open Quantum Safe Project for providing the liboqs library under an open-source licence.}

\conflictsofinterest{The authors declare no conflicts of interest.}

\abbreviations{Abbreviations}{
\noindent
\begin{tabular}{@{}ll}
AES & Advanced Encryption Standard\\
CRQC & Cryptographically Relevant Quantum Computer\\
ECDH & Elliptic-Curve Diffie--Hellman\\
FIPS & Federal Information Processing Standard\\
HKDF & HMAC-based Key Derivation Function\\
HNDL & Harvest Now, Decrypt Later\\
HVAC & Heating, Ventilation, and Air Conditioning\\
IoT & Internet of Things\\
KEM & Key Encapsulation Mechanism\\
MCU & Microcontroller Unit\\
ML-DSA & Module Lattice-Based Digital Signature Algorithm\\
ML-KEM & Module Lattice-Based Key Encapsulation Mechanism\\
MQTT & Message Queuing Telemetry Transport\\
NIST & National Institute of Standards and Technology\\
OQS & Open Quantum Safe\\
OTA & Over-the-Air\\
PQC & Post-Quantum Cryptography\\
RTT & Round-Trip Time\\
SRAM & Static Random Access Memory\\
TLS & Transport Layer Security
\end{tabular}
}

\begin{adjustwidth}{-\extralength}{0cm}
\bibliography{references}
\end{adjustwidth}
\end{document}